\documentclass[aps,amsmath,twocolumn]{revtex4}
\usepackage{graphicx}

\begin{document}
\title{Structure of a thermal quasifermion in the QCD/QED medium}
\author{Hisao Nakkagawa}%
\author{Hiroshi Yokota}%
\author{Koji Yoshida}%
\affiliation{%
 Institute for Natural Science, Nara University, 1500 Misasagi-cho, Nara 631-8502, Japan 
}%
\date{\today}
\begin{abstract}
In this paper we carried out a nonperturbative analysis of a thermal quasifermion in the chiral 
symmetric thermal QCD/QED medium by studying its self-energy function through the Dyson-Schwinger 
equation with the hard-thermal-loop resummed improved ladder kernel.

Our analysis reveals several interesting results, two in some of which may force us to change the image 
of thermal quasifermions: (1) The thermal mass of a quasifermion begins to decrease as the strength 
of the coupling gets stronger and finally disappears in the strong coupling region, thus showing a 
property of a massless particle, and (2) its imaginary part (i.e., the decay width) persists to 
have $O(g^2T \log(1/g))$ behavior. These results suggest that in the recently produced strongly coupled
quark-gluon plasma, 
the thermal mass of a quasiquark should vanish. Taking into account the largeness of the imaginary 
part, it seems very hard for a quark to exist as a qausiparticle in the  strongly coupled
quark-gluon plasma phase.

Other important findings are as follows: (3) The collective plasmino mode disappears also in the 
strongly coupled system, and (4) there exists an ultrasoft third peak in the quasifermion spectral 
density at least in the weakly coupled QED/QCD plasma, indicating the existence of the ultrasoft 
fermionic mode.
\end{abstract}
\maketitle

\section{\label{sec:intro}Introduction}
It is believed that the Relativistic Heavy Ion Collider (RHIC) at BNL and the Large Hadron 
Collider (LHC) at CERN have produced the primordial state of matter, namely, the quark-gluon 
plasma (QGP), and liberated the quark and gluon degrees of freedom. Subsequent analyses have 
shown that the produced QGP medium shows the property close to that of a perfect fluid. 
This fact leads us to the understanding that the QGP produced in the energy region of the RHIC 
and LHC is a strongly interacting system of quarks and gluons, namely, the strongly coupled 
QGP (sQGP)~\cite{review}.

Since the discovery of the sQGP phase, the behavior and the properties of the quasiquark in the new 
sQGP phase have attracted much attention; does the quark still work as the basic degree 
of freedom in the new phase or not ?  It is also pointed out theoretically that the 
hadronic excitation affects the spectral density of the quasiquark even in the chiral symmetric
phase, thus showing some characteristic structures near the phase 
boundary~\cite{Hatsuda-Kunihiro}. 
     
Up to now, most of the theoretical findings on thermal quasiquarks in the QGP are obtained 
through analyses with the assumption of weakly coupled QGP at high temperature, i.e., 
analyses through the hard-thermal-loop (HTL) resummed effective perturbation 
calculation~\cite{Rebhan}, or those through the one-loop calculation with the massive 
bosonic mode, or by replacing the thermal gluon with the massive vector boson~\cite{Kitazawa}. 
Kitazawa \textit{et al.}~\cite{Kitazawa} have pointed out the three-peak structure of the quasifermion spectral 
density and the existence of the massless third mode. Such analyses, however, cannot be 
justified in studying the thermal quasiparticle in the sQGP created in the energy region of 
RHIC. What we need is the nonperturbative analysis to explore the the properties of a strongly 
coupled system.

Nonperturbative calculations of correlators within lattice QCD are performed in Euclidean 
space and give interesting results~\cite{Schaefer}. However, strictly speaking it is not possible 
to carry out an analytic continuation that is necessary to determine the spectral function. 
In addition, it is difficult on the lattice to respect the chiral symmetry that should be 
restored in the sQGP phase, though we are interested in the property of thermal quasiparticles 
in the chiral symmetric sQGP phase.

In this paper we perform a nonperturbative analysis of a thermal quasifermion in thermal 
QCD/QED by studying its self-energy function through the Dyson-Schwinger equation (DSE) 
with the HTL resummed improved ladder kernel. Our analysis may overcome the problems in 
the previous analyses listed above for the following reasons:  (1) it is a nonperturbative 
QCD/QED analysis, (2) we study the DSE in the real-time formalism of thermal field theory, 
which is suitable for the direct calculation of the propagator, or the spectral function, 
(3) we use the HTL resummed thermal gauge boson (gluon/photon) propagator as an interaction 
kernel of the DSE, and take into account the quasiparticle decay processes by accurately 
studying the imaginary part of the self-energy function, and finally, (4) we present an 
analysis based on the DSE that respects the chiral symmetry and describes its dynamical 
breaking and restoration. Our analysis is nothing but an application of our formalism 
employing the DSE to the study of thermal quasifermions on the strongly coupled QCD/QED 
medium with chiral symmetry~\cite{FNYY2,NYY_proc}.

With the solution of the DSE with the HTL resummed improved ladder kernel, 
we study the properties of the thermal quasifermion spectral density and its peak 
structure,as well as the dispersion law of the physical modes corresponding to the poles of thermal 
quasifermion propagator, through which we elucidate the properties of the thermal mass and the 
decay width of fermion and plasmino modes, and also pay attention to properties of the 
possible third mode, especially in the sQGP phase.

Analogous studies employing the DSE are carried out by several groups~\cite{Harada}. 
All these analyses solve the DSE in the imaginary-time formalism, and try to 
perform an analytic continuation. Harada \textit{et al.} study the DSE with a ladder kernel 
in which the tree-level gauge boson propagator is used, 
while Qin \textit{et al.} and Mueller \textit{et al.} use the maximum entropy method to 
compute the quark spectral density. Qin \textit{et al.} also pays a special attention 
to the massless third mode.

Our analysis reveals several interesting results, two in some of which may force 
us to change the image of thermal quasifermions: (1) The thermal mass of a 
quasifermion begins to decrease as the strength of the coupling gets stronger 
and finally disappears in the strong coupling region, thus showing a property of 
a massless particle, and (2) its imaginary part (i.e., the decay width) persists 
to have $O(g^2T \log(1/g))$ behavior. These results suggest that in the recently 
produced sQGP, the thermal mass of a quasiquark should vanish. Taking into account 
the largeness of the imaginary part (i.e., the decay width), it seems very hard 
for a quark to exist as a qausiparticle in the sQGP phase.

Other important findings are as follows: (3) The collective plasmino mode disappears 
also in the strongly coupled system, and (4) there exists an ultrasoft third peak 
in the quasifermion spectral density at least in the weakly coupled QED/QCD plasma, 
indicating the existence of the ultrasoft fermionic mode.

Focusing on fact (1) above, we have already reported briefly on this in Ref.~\cite{NYY11}, 
and in the present paper we give results of a detailed analysis. 
Fact (4) has also been pointed out briefly in Ref.~\cite{NYY11}, and will be studied in
further detail in a separate paper.

This paper is organized as follows; In Sec.~\ref{sec:DSE} we present the HTL resummed 
improved ladder  Dyson-Schwinger equation for the quasifermion self-energy function, 
with which we investigate the property of the thermal quasifermion in the chirally symmetric 
QGP phase, and give the results in Sec.~\ref{sec:Solution}. 
In Sec.~\ref{sec:Solution-Density}, properties of the 
quasifermion spectral density are studied, and in Sec.~\ref{sec:Solution-Peak} 
we discuss the problem in the 
relation between the peak position of spectral density and the zero point of the inverse 
quasifermion propagator. The dispersion law of the quasifermion is studied in 
Sec.~\ref{sec:Solution-Dispersion}
and the vanishment of thermal mass and the disappearance of 
the plasmino mode in the strongly coupled system are pointed out. 
Properties of the thermal mass and the existence of the third peak or the ultrasoft 
mode are discussed in Secs.~\ref{sec:Solution-Thermal_mass}
and \ref{sec:Solution-3rdPeak}, respectively. 
Finally in Sec.~\ref{sec:Solution-DecayWidth}, properties of the decay width of quasifermion  
are studied. A summary of the paper and discussion are given in Sec.~\ref{sec:Discussion}. 
Several appendixes are also given.  
In Appendix~\ref{ap:HTL-DSE} we explain the approximations to get the HTL resummed 
improved ladder DSE to be solved. The cutoff dependence of our analysis 
is discussed in Appendix~\ref{ap:cutoff-dep}, and the phase boundary between the chirally 
symmetric and broken phases in the Landau gauge is briefly explained 
in Appendix~\ref{ap:Phase}. Finally Appendix~\ref{ap:Disp-Law-rho} 
is devoted to explaining why we do not use 
the peak position of the spectral density as the condition to determine 
the on-shell particle.

\section{\label{sec:DSE}The HTL resummed improved ladder Dyson-Schwinger equation}

In this paper we study the thermal QCD/QED in the real-time closed time-path 
formalism~\cite{CTP-review}, 
and solve the DSE for the retarded fermion self-energy 
function $\Sigma_R$, to investigate the property of the thermal quasifermion in the chiral 
symmetric QGP phase. Throughout this paper we study the massless QCD/QED in the 
Landau gauge.

As is well-known, at zero temperature the Landau gauge plays an essential role to 
ensure the gauge invariance of the solution of the ladder DSE 
because it is proved that in the Landau gauge the 
wave function receives no renormalization, i.e., $A(P)=1$~\cite{Mas-Naka,Kugo-Naka}.  
At finite temperature, however, $A(P) \neq 1$ even in the 
Landau gauge, and there is no special reason to choose the Landau gauge any more.
 In this analysis we choose the Landau gauge first for the sake of simplicity, and 
second for the sake of comparison with other works.

In this section we present the HTL resummed DSE for $\Sigma_R$, and also give an 
explication about the improved ladder approximation we make use of to the HTL 
resummed gauge boson propagator.  We also calculate the effective potential for the 
retarded fermion propagator $S_R$ in order to find the ``true solution'' when we get several 
``solutions'' of the DSE. 

\subsection{\label{sec:DSE-Sigma}HTL resummed improved ladder DSE for fermion self-energy
function $\Sigma_R$}
	
The retarded quasifermion propagator $S_R(P), P=(p_0, \text{\bf p})$, is expressed by
\begin{equation}
\label{eq_SR}
   S_R(P) = \frac{1}{P\!\!\!\!/ + i \epsilon \gamma^0 - \Sigma_R (P)} .
\end{equation}
The retarded fermion self-energy function $\Sigma_R$ can be tensor-decomposed in a chiral
symmetric phase at finite temperature as follows:
\begin{equation}
\label{eq_Sigma}
   \Sigma_R(P) = (1-A(P)) p_i \gamma^i - B(P) \gamma^0 .
\end{equation}
$A(P)$ is the inverse of the fermion wave function renormalization function, and $B(P)$ is the 
chiral invariant mass function. The $c$-number mass function does not appear in the chiral symmetric phase.

In the real-time closed time-path formalism, by adopting the tree vertex and the HTL resummed gauge
 boson propagator for the interaction kernel of the DSE, we obtain, in the massless thermal QED/QCD, 
the HTL resummed improved ladder DSE for retarded fermion self-energy function $\Sigma_R$~\cite{FNYY1,FNYY2} 
(coupling $\alpha \equiv g^2/4\pi$ : $g^2 = g_s^2 C_f$ for QCD, $g= e$ for QED) %
\begin{eqnarray}
\!\!\!\! & &- i \Sigma_R(P) = - \frac{g^2}{2} \int \frac{d^4K}{(2 \pi)^4} \nonumber \\
 & & \ \ \times \left[ {}^* \Gamma^{\mu}_{RAA}(-P,K,P-K) S_{RA}(-K,K) \right. \nonumber \\
 & & \ \ \ \ \times {}^* \Gamma^{\nu}_{RAA} (-K,P,K-P) {}^*G_{RR,\mu \nu}(K-P,P-K)
             \nonumber \\
 & & \ \  + {}^* \Gamma^{\mu}_{RAA}(-P,K,P-K) S_{RR}(-K,K) \nonumber \\
 & & \ \ \ \ \left. \times {}^* \Gamma^{\nu}_{AAR} (-K,P,K-P) {}^*G_{RA,\mu \nu}(K-P,P-K) 
              \right] .   \nonumber \\
\label{eq_DSE}
\end{eqnarray}
Here $^*G_{\mu\nu}$ is the HTL resummed gauge boson propagator~\cite{Klimov,Weldon} 
where  $R \equiv RA$ and $C \equiv RR$ denote the retarded and the correlation components, respectively, 
and $^*\Gamma_{\mu} = \gamma_{\mu}$ in the present approximation.

There have been many attempts to carry out the higher order calculation within the HTL resummed effective 
perturbation theory and to get information beyond the applicability region of the HTL 
approximation~\cite{Rebhan,Bra-Nieto,Anderson}.  
The DSE with the HTL 
resummed gauge boson propagator as an interaction kernel can take the dominant effects of 
thermal fluctuation of $O(gT)$ into account nonperturbatively. Thus we expect the HTL 
resummed improved ladder DSE to enable us to study wider regions of the couplings and 
temperatures, e.g., the strongly coupled QGP medium, than those restricted by the HTL 
approximation, i.e., the regions of weak couplings and high temperatures.

The explicit expression of the HTL resummed improved ladder DSE in the Landau gauge to determine 
the scalar invariants $A(P)$ and $B(P)$ in the chiral symmetric QGP phase becomes coupled integral 
equations as follows:%
\begin{widetext}
\begin{subequations}
\label{eq_DSE_AB}
\begin{eqnarray}
\label{eq_DSE_A}
 p^2[1-A(P)] &=& g^2 \left. \int \frac{d^4K}{(2 \pi)^4}
       \right[ \{1+2n_B(p_0-k_0) \} \text{Im}[\ ^*G^{\rho \sigma}_R(P-K)]
       \times  \nonumber \\
  & & \Bigl[ \{ K_{\sigma}P_{\rho} + K_{\rho} P_{\sigma}
       - p_0 (K_{\sigma} g_{\rho 0} + K_{\rho} g_{\sigma 0} ) 
       - k_0 (P_{\sigma} g_{\rho 0} + P_{\rho} g_{\sigma 0} )
       + pkz g_{\sigma \rho} \nonumber \\
  & & + 2p_0k_0g_{\sigma 0}g_{\rho 0} \}\frac{A(K)}{[k_0+B(K)+i
       \epsilon]^2 - A(K)^2k^2 }
       + \{ P_{\sigma} g_{\rho 0} + P_{\rho} g_{\sigma 0} \nonumber \\
  & & - 2p_0 g_{\sigma 0} g_{\rho 0} \}
       \frac{k_0+B(K)}{[k_0+B(K)+i \epsilon]^2 - A(K)^2k^2
       } \Bigr] + \{1-2n_F(k_0) \}
       \times \nonumber \\ 
  & & \ ^*G^{\rho \sigma}_R(P-K) \text{Im} \Bigl[
       \{ K_{\sigma}P_{\rho}  + K_{\rho} P_{\sigma} - p_0 (K_{\sigma}
       g_{\rho 0} + K_{\rho} g_{\sigma 0} ) - k_0 (P_{\sigma}
       g_{\rho 0} + P_{\rho} g_{\sigma 0} ) \nonumber \\
  & & + pkz g_{\sigma \rho} + 2p_0k_0g_{\sigma 0}g_{\rho 0}\}
       \frac{A(K)}{[k_0+B(K)+i \epsilon]^2 - A(K)^2k^2} 
       \nonumber \\
  & & \left. +  \{ P_{\sigma} g_{\rho 0} + P_{\rho} g_{\sigma 0}
       - 2p_0 g_{\sigma 0} g_{\rho 0} \}
       \frac{k_0+B(K)}{[k_0+B(K)+i \epsilon]^2 - A(K)^2k^2 } \Bigr] \right] \ , \\
\label{eq_DSE_B}
 B(P) &=& g^2 \left. \int \frac{d^4K}{(2 \pi)^4} \right[
        \{1+2n_B(p_0-k_0)\} \text{Im}[\ ^*G^{\rho \sigma}_R(P-K)] \times
         \nonumber \\
  & & \Bigl[ \{ K_{\sigma} g_{\rho 0} + K_{\rho} g_{\sigma 0}
       - 2k_0 g_{\sigma 0} g_{\rho 0} \}
       \frac{A(K)}{[k_0+B(K)+i \epsilon]^2 - A(K)^2k^2} \nonumber \\
  & & + \{ 2g_{\rho 0} 2g_{\sigma 0} - g_{\sigma \rho} \} 
       \frac{k_0+B(K)}{[k_0+B(K)+i \epsilon]^2 - A(K)^2k^2 }
       \Bigr] + \{1-2n_F(k_0) \} \times \nonumber \\ 
  & & \ ^*G^{\rho \sigma}_R(P-K) \text{Im} \Bigl[ \frac{A(K)}{[k_0+B(K)+i
       \epsilon]^2 - A(K)^2k^2 } 
       \{ K_{\sigma} g_{\rho 0} + K_{\rho} g_{\sigma 0}  \nonumber \\
  & & \left. - 2k_0 g_{\sigma 0} g_{\rho 0} \} + \frac{k_0+B(K)}{[k_0+B(K)+
       i \epsilon]^2 - A(K)^2k^2}
       \{ 2g_{\rho 0} 2g_{\sigma 0} - g_{\sigma \rho} \} \Bigr] 
       \right] \ , 
\end{eqnarray}
\end{subequations}
where $n_B(x)$ and $n_F(x)$ are the Bose-Einstein and the Fermi-Dirac equilibrium distribution 
functions, respectively,
\begin{equation}
\label{eq_NB_BF}
 n_B(x) = \frac{1}{\exp(x/T)-1}  ,  \ \ \ \ \ \ \ \ \ \ \  
 n_F(x) = \frac{1}{\exp(x/T) +1}.  \nonumber
\end{equation}
\end{widetext}

The above DSEs, Eq.~(\ref{eq_DSE_AB}), are still very tough to attack, and need
further approximations to be solved. We thus adopt the instantaneous exchange 
approximation to the longitudinal gauge boson propagator; i.e., we set the zeroth component of 
the longitudinal gauge boson momentum $q_0$ to zero. Details of the approximation we use are 
explained in Appendix~\ref{ap:HTL-DSE}.

In solving the DSEs, Eq.~(\ref{eq_DSE_AB}), we are forced to introduce a momentum 
cutoff in the 
integration over the four-momentum $\int d^4K$, $K=(k_0, \text{\bf k})$. 
We use the following cutoff method  ($\Lambda$ denotes an arbitrary cutoff parameter 
and plays a role to scale any dimensionful quantity, e.g., $T = 0.3$ means $T = 0.3\Lambda$):
\begin{center}
\begin{tabular}{llcl}
three-momentum & ~$\text{\bf k}$~ & ~:~ & $k=|\text{\bf k}| \le \Lambda$,  \\
energy         & ~$k_0$~ & ~:~ & $|k_0| \le  \Lambda_0$ \  ($\Lambda_0=\Lambda \sim 5\Lambda$). \\
\end{tabular}
\end{center}
We determine the parameter $\Lambda_0$ so as to get a stable solution for the fermion spectral density. 
Over the range of the temperature and the coupling we study in the present analysis,
we set $\Lambda_0=2\Lambda$. In fact the solution is totally stable for $\Lambda_0 \ge 2\Lambda$.
In Appendix~\ref{ap:cutoff-dep} we give details of the cutoff dependence of the solution.

\subsection{\label{sec:DSE-potential}The effective potential $V[S_R]$ for the retarded full fermion propagator $S_R$}

The above DSEs, Eq.~(\ref{eq_DSE_AB}), may have several solutions, and we choose the ``true'' solution by 
evaluating the effective potential $V[S_R]$ for the fermion propagator function $S_R$, then finding
 the lowest energy solution. The effective potential is expressed 
as~\cite{CJT}
\begin{widetext}
\begin{eqnarray}
\label{eq_potential}
 V [ S_R ] \!\! & =& \!\! \text{Tr} \left[ P\!\!\!\!/ S_R \right] +  \text{Tr} \ln  \left[ S_R^{-1} \right]   + \frac{g^2}{2} i \int \frac{d^4K}{(2 \pi)^4} \int \frac{d^4P}{(2 \pi)^4}
             \frac12 \text{tr} \left[ \gamma_{\mu} S_R(K) \gamma_{\nu} S_R(P) G_C^{\mu \nu} (P-K) \right. \nonumber \\
       & & \ \ \ \ \ \ \left.
               +  \gamma_{\mu} S_C(K) \gamma_{\nu} S_R(P) G_R^{\mu \nu} (P-K)
               +  \gamma_{\mu} S_R(K) \gamma_{\nu} S_C(P) G_A^{\mu \nu} (P-K) \right] , 
\end{eqnarray}
\end{widetext}
where the first and the second terms correspond to the one-loop effective potential, while the third term 
corresponds to the two-loop contribution.

\section{\label{sec:Solution}Solution of the HTL resummed improved ladder Dyson-Schwinger equation}

In this section we give the solution of the HTL resummed improved ladder DSE for the retarded 
self-energy function $\Sigma_R$, and study its consequences in the chiral symmetric phase. 
Special attention is paid to the consequences in the strongly coupled QCD/QED medium,
in order to get information on the thermal quasifermion in the strongly coupled QGP (sQGP) phase, 
recently discovered through the experiments at RHIC and LHC~\cite{review}.
Part of the results are already briefly reported~\cite{NYY11}.  

\subsection{\label{sec:Solution-Density}Quasifermion spectral density}

\subsubsection{\label{sec:Solution-Density-rho}Spectral density of quasifermion $\rho_{\pm}$}

In the chiral symmetric QCD/QED phase, the quasifermion propagator can be expressed as
\begin{equation}
\label{eq_SR_fact}
S_R (P) = \frac12 \left[ \frac{1}{D_+} \left( \gamma^0 + \frac{p_i \gamma^i}{p} \right)
          + \frac{1}{D_-} \left( \gamma^0 - \frac{p_i \gamma^i}{p} \right) \right] 
\end{equation}
where
\begin{equation}
\label{eq_D_def}
        D_{\pm}(P)=p_0+B(p_0,p) \mp p A(p_0,p)
\end{equation}
with
\begin{subequations}
\label{eq_D_sym}
\begin{eqnarray}
\label{eq_ReD_sym}
             \text{Re} [D_+(p_0, p)] &=&  - \text{Re} [D_-(-p_0, p)], \\
\label{eq_ImD_sym}
             \text{Im} [D_+(p_0, p)] &=&   \text{Im} [D_-(-p_0, p)].
\end{eqnarray}
\end{subequations}

The spectral density of quasifermion $\rho_{\pm}$ is defined by
\begin{eqnarray}
  \rho_{\pm}(p_0,p) &=& - \frac{1}{\pi} \text{Im} \frac{1}{D_{\pm}(P)} \nonumber \\
         &=& - \frac{1}{\pi} \text{Im} \frac{1}{p_0 + B(p_0,p) \mp p A(p_0,p)},
\label{eq_rho_def}
\end{eqnarray}
which satisfies the sum rules~\cite{LeBellac}
\begin{subequations}
\label{eq_sum_rules}
\begin{eqnarray}
\label{eq_sum_rule_1}
  \int^{\infty}_{-\infty} dp_0 \rho_{\pm}(p_0,p) &=& 1 ,\\
\label{eq_sum_rule_2}
  \int^{\infty}_{-\infty} dp_0 p_0 \rho_{\pm}(p_0,p) &=& \pm p, \\
\label{eq_sum_rule_3}
  \int^{\infty}_{-\infty} dp_0 p_0^2 \rho_{\pm}(p_0,p) &=& p^2+m^2_f .
\end{eqnarray}
\end{subequations}
It also satisfies the symmetry property
\begin{equation}
  \rho_{\pm}(p_0,p) = \rho_{\mp}(-p_0,p) .
\label{eq_rho_sym}
\end{equation}
We therefore study only $\rho_+(p_0, p)$ throughout this paper. 

To study the spectral properties of the quasifermion in the chiral symmetric 
QGP phase, we must at first make sure that we are in fact studying inside 
the chiral symmetric phase. We have already studied the phase structure of 
thermal QCD/QED through the same DSE approach~\cite{FNYY2}, and determined the 
phase boundary between the chiral symmetric and the chiral symmetry broken 
phases. Thus we are sure that the region of temperatures and couplings we study 
in the present paper are well within the chiral symmetric phase. For details, 
see Appendix~\ref{ap:Phase}, where we give the results of our analyses to determine the phase boundary. 

It is to be noted that the temperature $T$ in the present analysis represents the temperature scaled 
by the momentum cutoff $\Lambda$, thus $0 \le T \le 1$, and it does not denote the real temperature
itself. With the temperature thus defined, we can determine the phase boundary curve in the 
two-dimensional ($\alpha, T$) plane, separating the chiral symmetry broken/restored phase~\cite{FNYY2}. 
If we measure the temperature $T$ relative to the critical temperature $T_c$, high or low temperature 
has a definite meaning

\subsubsection{\label{sec:Solution-Density-Structure}Structure of quasifermion spectral density}

\begin{figure*}[htbp]
\begin{tabular}{ccc}
\begin{minipage}{0.33\hsize}
  \centerline{\includegraphics[width=7.5cm]{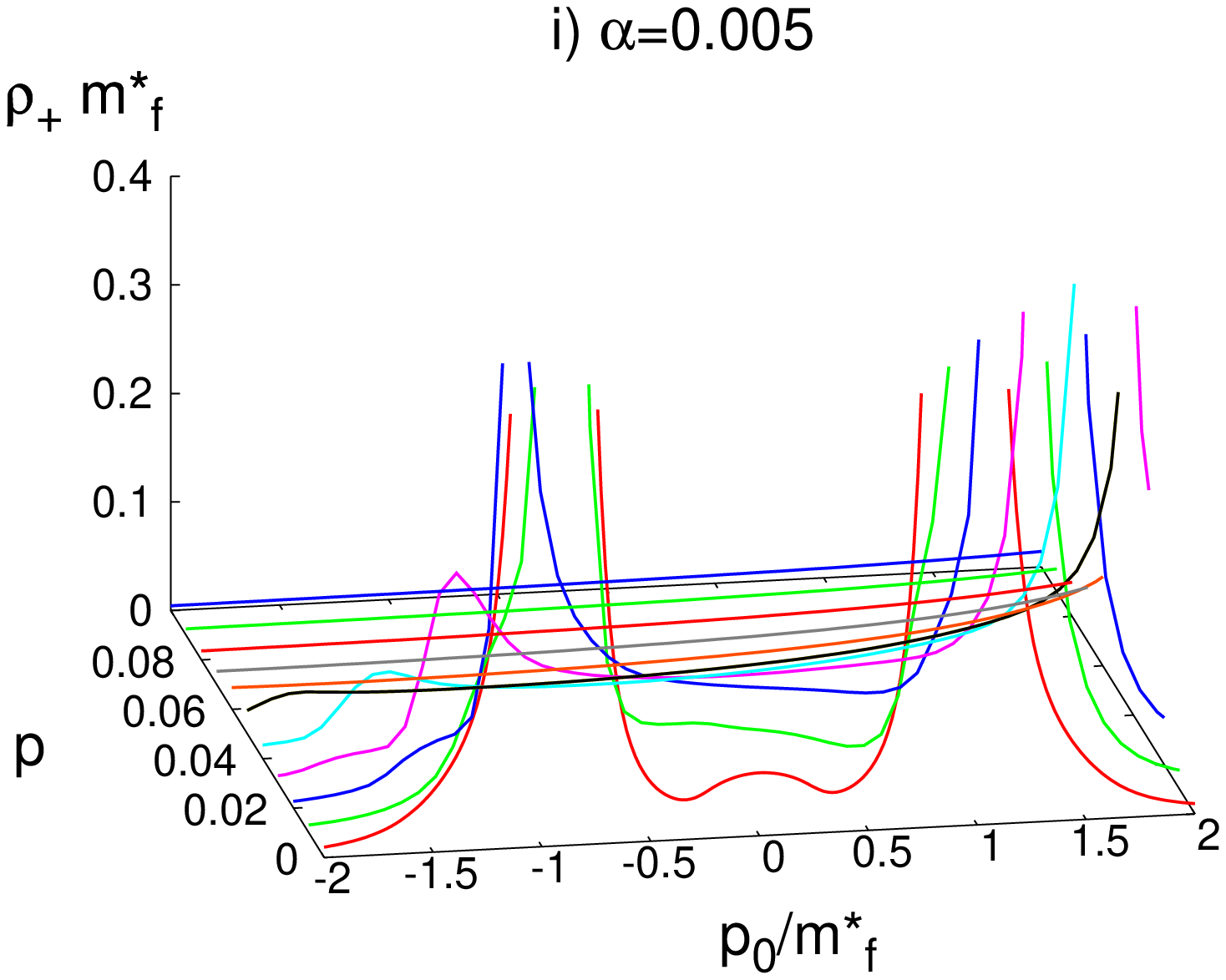}}
\end{minipage}
\begin{minipage}{0.33\hsize}
  \centerline{\includegraphics[width=7.5cm]{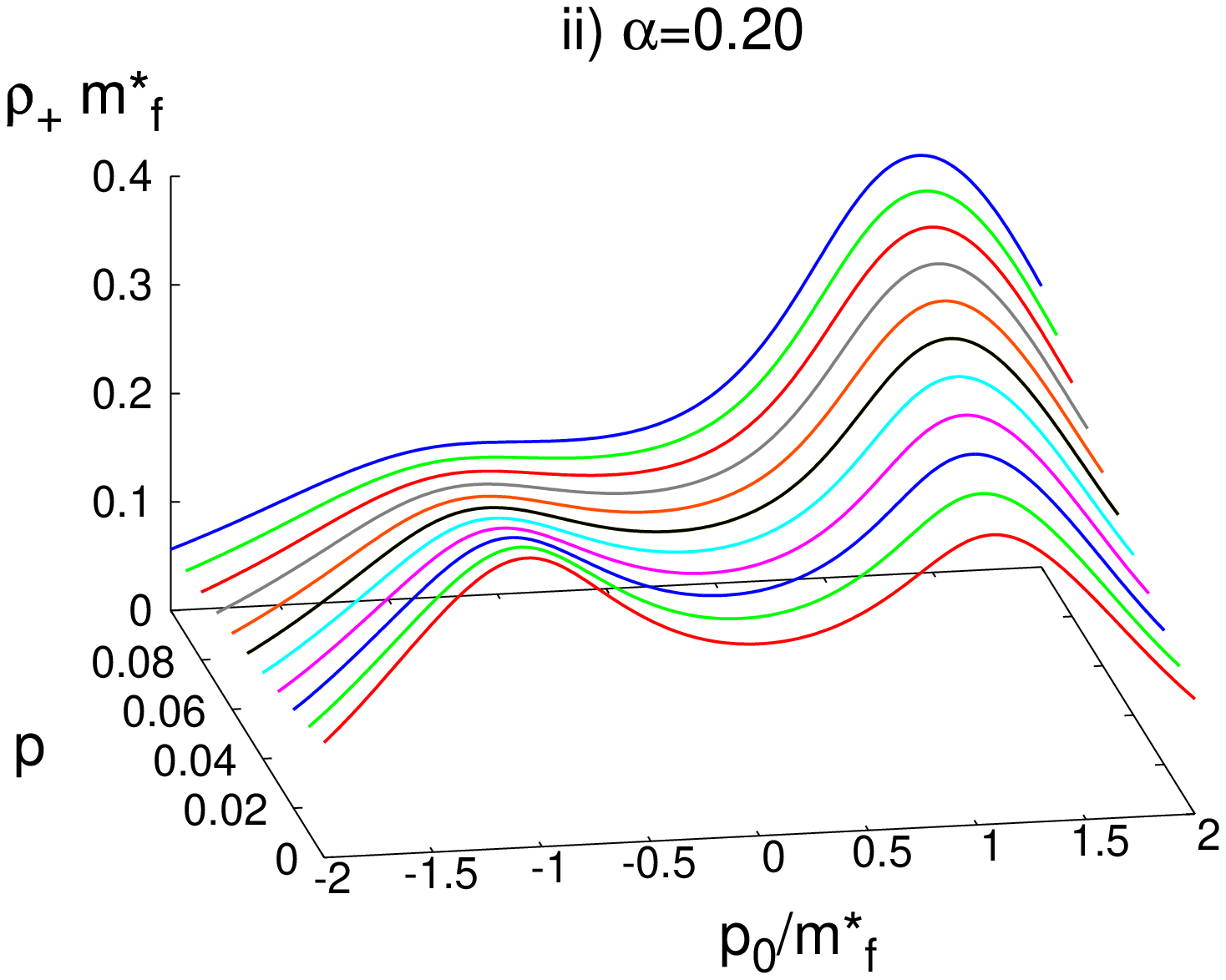}}
\end{minipage}
\begin{minipage}{0.33\hsize}
  \centerline{\includegraphics[width=7.5cm]{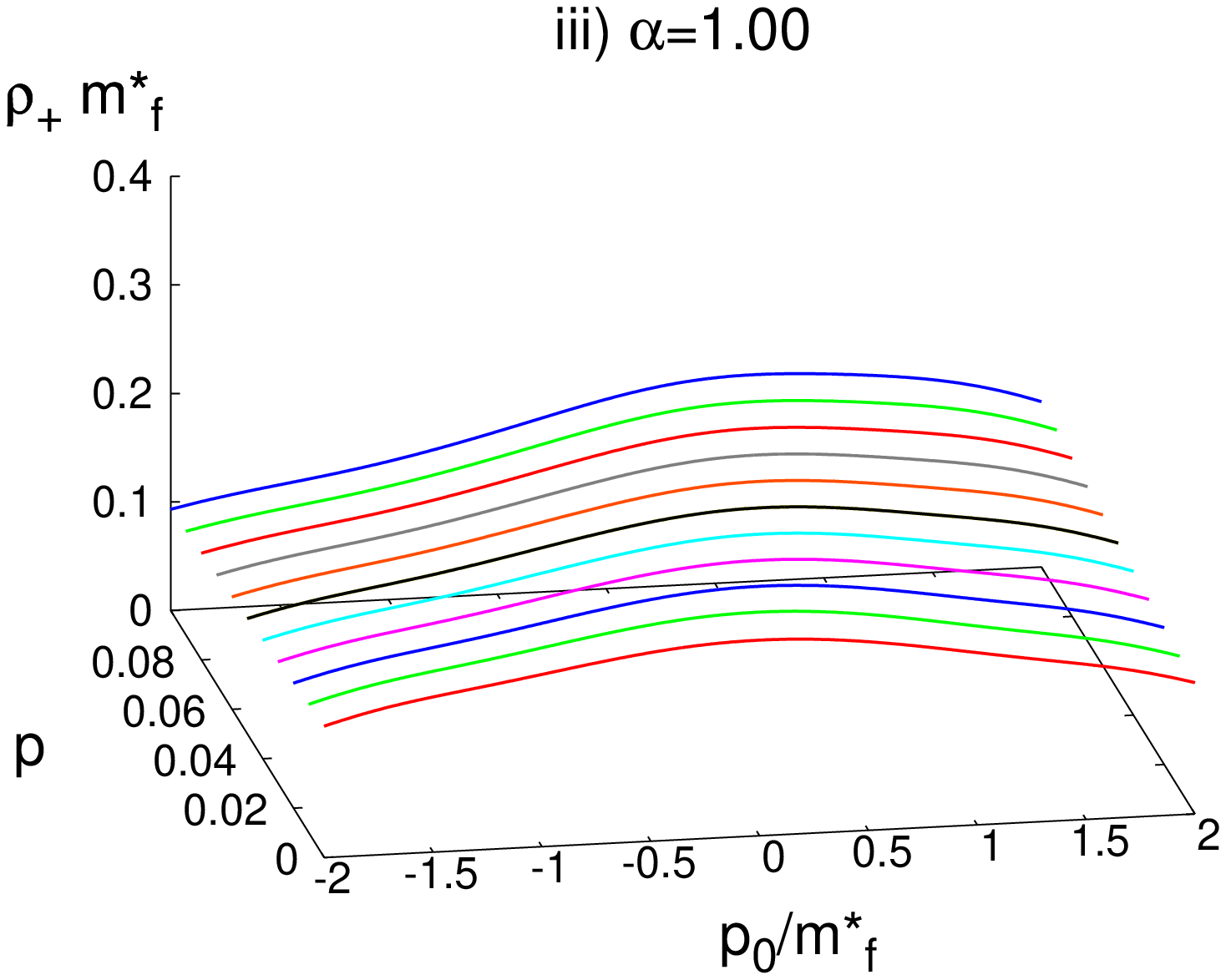}}
\end{minipage}
\end{tabular}
\caption{The coupling $\alpha$ dependence of the quasifermion spectral density $\rho_+ (p_0,p)$ at $T=0.3$.}
\label{3d_rho}
\end{figure*}  

\begin{figure*}[htbp]
\begin{tabular}{ccc}
\begin{minipage}{0.50\hsize}
  \centerline{\includegraphics[width=7.5cm]{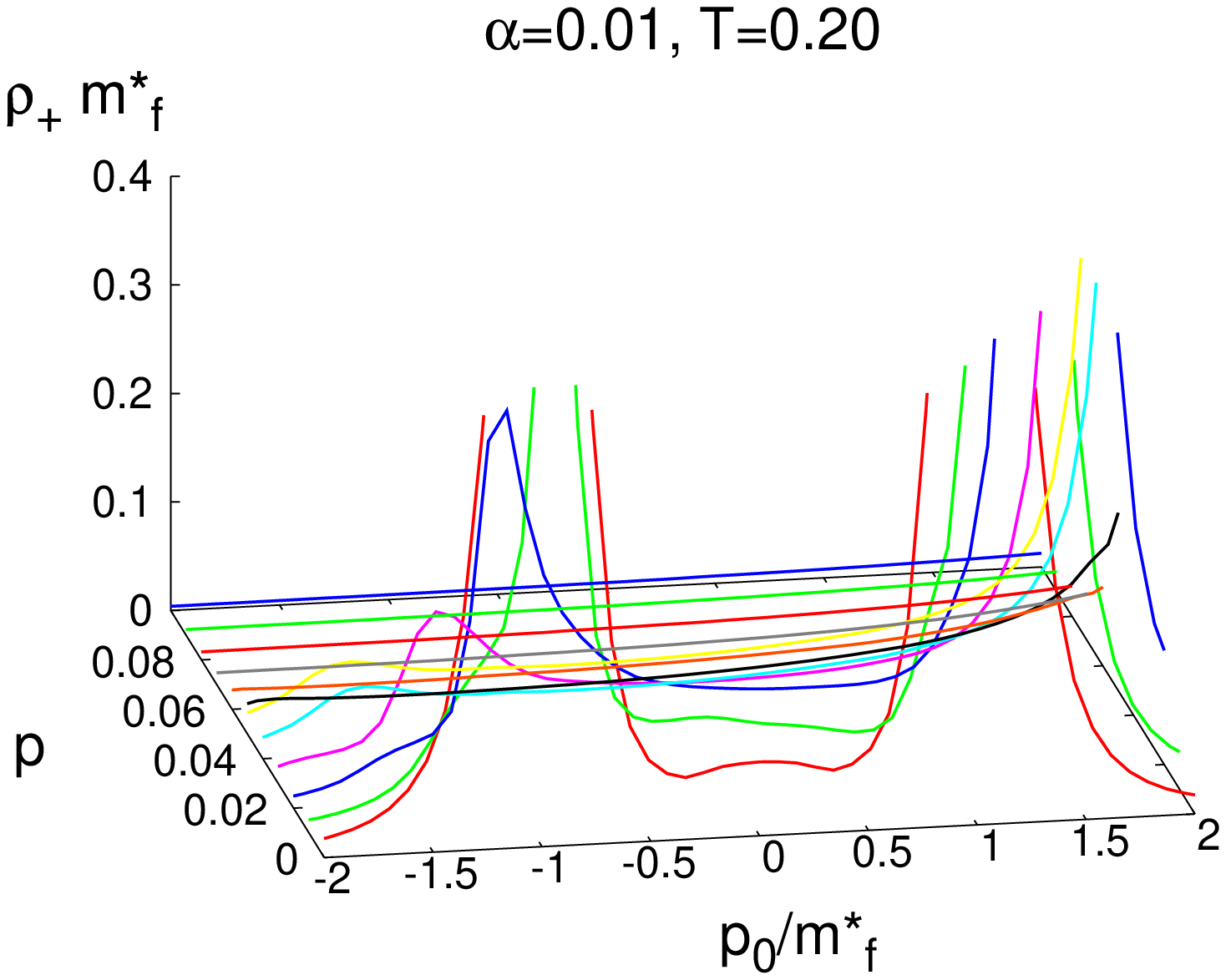}}
\end{minipage}
\begin{minipage}{0.50\hsize}
  \centerline{\includegraphics[width=7.5cm]{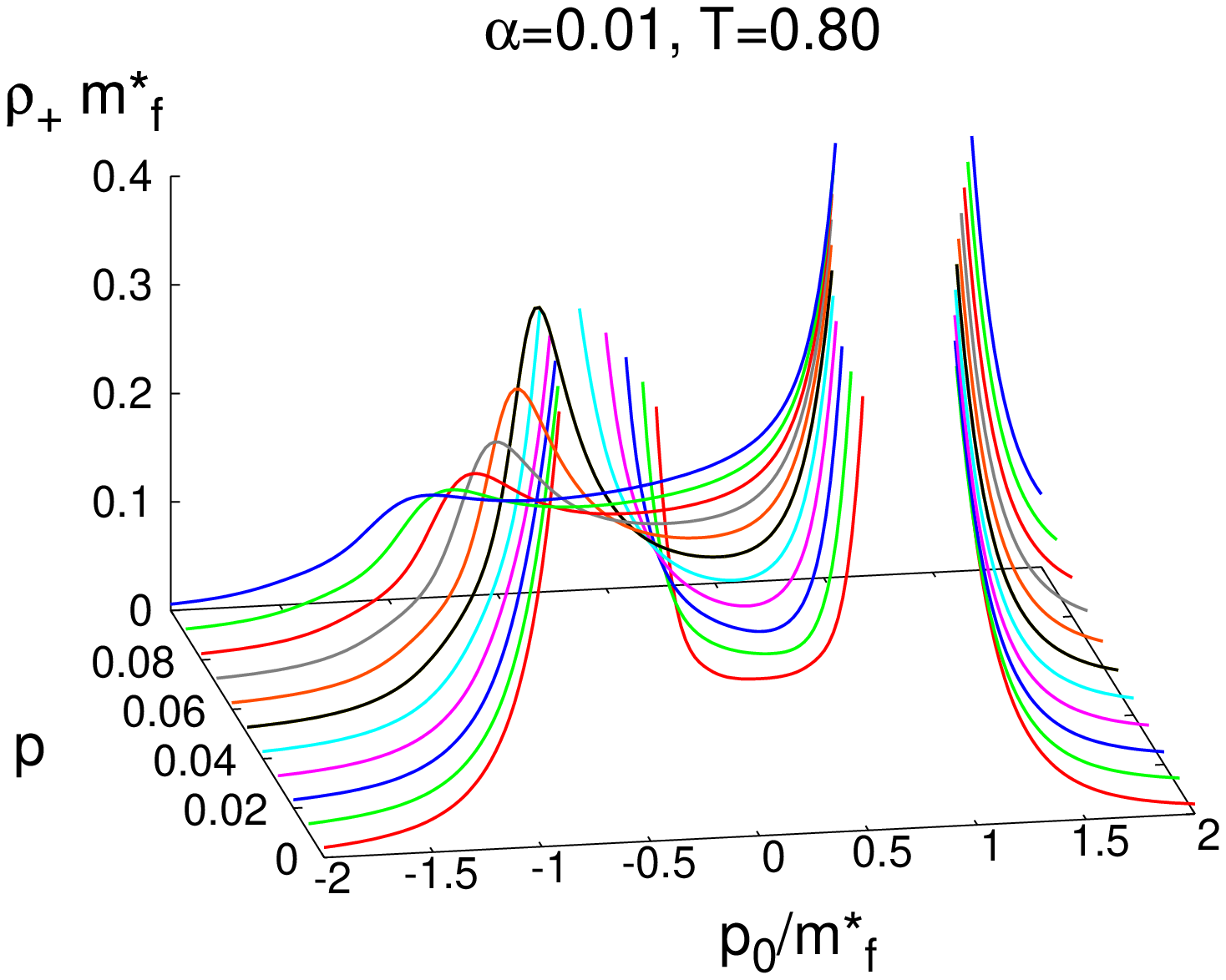}}
\end{minipage} \\
\begin{minipage}{0.50\hsize}
  \centerline{\includegraphics[width=7.5cm]{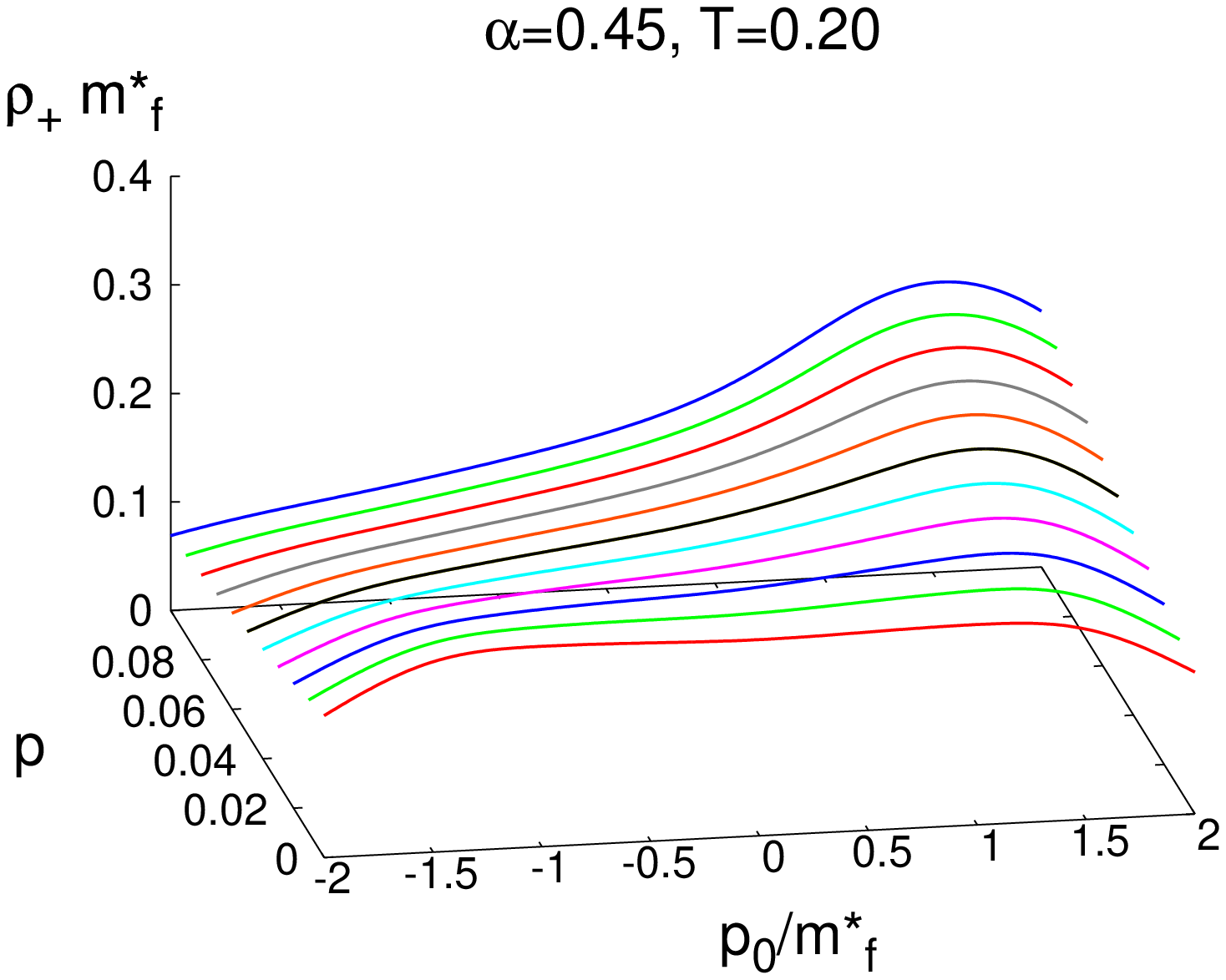}}
\end{minipage}
\begin{minipage}{0.50\hsize}
  \centerline{\includegraphics[width=7.5cm]{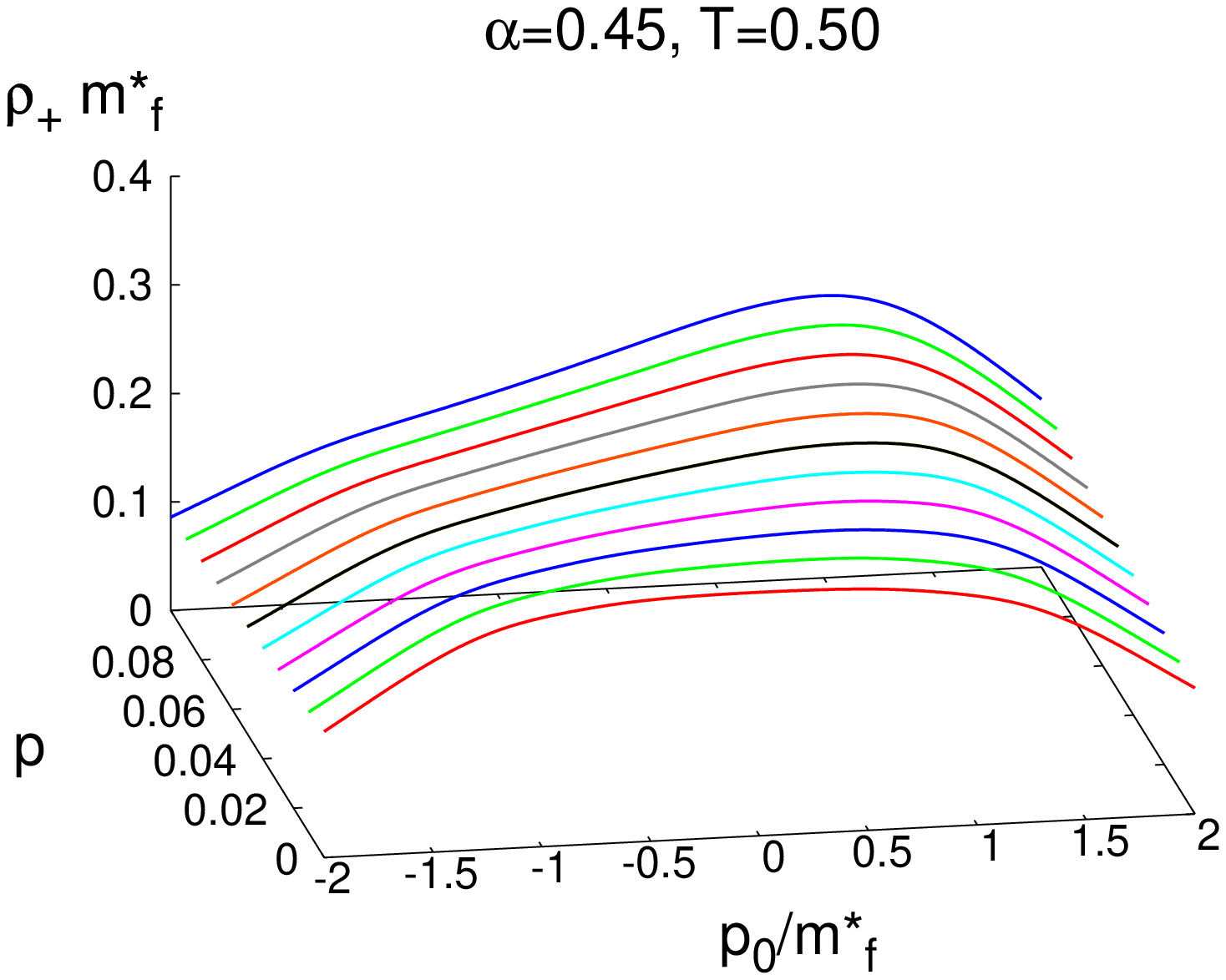}}
\end{minipage}
\end{tabular}
\caption{The temperature $T$ dependence of the quasifermion spectral density $\rho_+(p_0,p)$ 
at the weak coupling $\alpha=0.01$ (upper two graphs)
and at the  strong coupling $\alpha=0.45$ (lower two graphs).}
\label{3d_rho_afix}
\end{figure*}  

Now let us present the properties of quasifermion spectral density $\rho_+$ computed by the 
solution of DSEs, Eq.~(\ref{eq_DSE_AB}). 
At first we should note the fact that the quasifermion spectral density $\rho_+$ thus
determined well satisfies the sum rules Eqs.~(\ref{eq_sum_rule_1}) and (\ref{eq_sum_rule_2})
within a few percent error. This fact proves \textit{a posteriori} the adequacy of our choice
of  cutoff parameter $\Lambda_0=2\Lambda$ in the region of the couplings and the temperatures we 
study.  The third sum rule is heavily dependent on the HTL calculation, and the agreement
depends on the couplings and the temperatures.

Next let us show the structure of $\rho_+(p_0, p)$ as a 
function of $p_0$ and $p$ in the two-dimensional $(p_0, p)$ plane. 
For convenience we study the dimensionless quantity $\rho_+(p_0,p) m^*_f$,
where $m^*_f$ denotes the thermal mass determined through the next-to-leading
order calculation of the HTL resummed effective perturbation
theory~\cite{Rebhan,Schulz},
\begin{eqnarray}
\label{eq_mf_star}
     \left( \frac{m^*_f}{m_f} \right)^2 &=& 1 - \frac{4 g}{\pi} \left[
            - \frac{g}{2 \pi} + \sqrt{ \frac{g^2}{4\pi^2} + \frac13} \right] , \\
         m_f^2  &\equiv& \frac{g^2T^2}{8} . \nonumber
\end{eqnarray}

In measuring at moderately high temperature $T=0.4$, we can see in Fig.~\ref{3d_rho} 
the three typical 
peak structures depending on the strength of the coupling $\alpha=g^2/4\pi$:

\begin{description}
\item{(i)} \ \ At weak coupling $\alpha=0.005$ we can see three peaks as a function of $p_0$ at $p=0$.
Two sharp peaks of them at positive and negative $p_0$ represent the fermion and the collective 
plasmino modes, respectively~\cite{Klimov,Weldon},
and the slight third ``peak'' barely recognizable around $p_0=0$ 
corresponds to the massless, or the ultrasoft mode~\cite{Kitazawa,Hidaka}. 
The plasmino mode and 
the massless mode rapidly decrease and disappear as the size of momentum $p$ becomes large. 
\item{(ii)} \ \ At the intermediate strength $\alpha=0.2$, we can see only two peaks at $p_0 \neq 0$ 
as a function of $p_0$ even at $p=0$, and unable to recognize the existence of the third peak corresponding 
to the massless pole in this region of the coupling. The peak at the negative side of the $p_0$ axis that
may correspond to the collective plasmino pole rapidly disappears as $p$ gets large. 
\item{(iii)} \ \ At the strong coupling $\alpha=1.0$ we can only recognize, at any size of the momentum $p$, 
the existence of a broad ``peak'' that may represent the massless pole. No massive pole exists 
in the strong coupling region. 
\end{description}

Note the vast differences of the height of the peak and the spread of the spectral 
density in the three cases of the coupling strength (i), (ii), and (iii), clearly showing the broadness 
of the ``peak'' in the strong coupling environment. 

Figure~\ref{3d_rho} shows the coupling $\alpha$ dependence of the quasifermion spectral 
density $\rho_+ m^*_f$ at fixed temperature $T=0.3$. We can also see the temperature $T$ dependence, 
which is given in Fig.~\ref{3d_rho_afix}.

In Fig.~\ref{3d_rho}, as noted above, we see the transition of the peak structure of spectral density as 
the strength of the coupling varies with the temperature kept fixed: triple peaks at small couplings, double 
peaks at intermediate couplings and finally single peak at strong couplings. 

Figure~\ref{3d_rho_afix} shows that the analogous behavior is also observed when the temperature of 
the environment varies with the strength of the coupling kept fixed. At small couplings ($\alpha=0.01$), 
the three-peak structure at low temperature ($T=0.2$) changes to the double-peak structure at high 
temperature ($T=0.8$), and at intermediate couplings ($\alpha=0.45$), the double-peak structure at low 
temperature ($T=0.2$) tends to the single-peak structure at high temperature ($T=0.8$). 

\begin{figure}[htbp] 
  \centerline{\includegraphics[width=7.5cm]{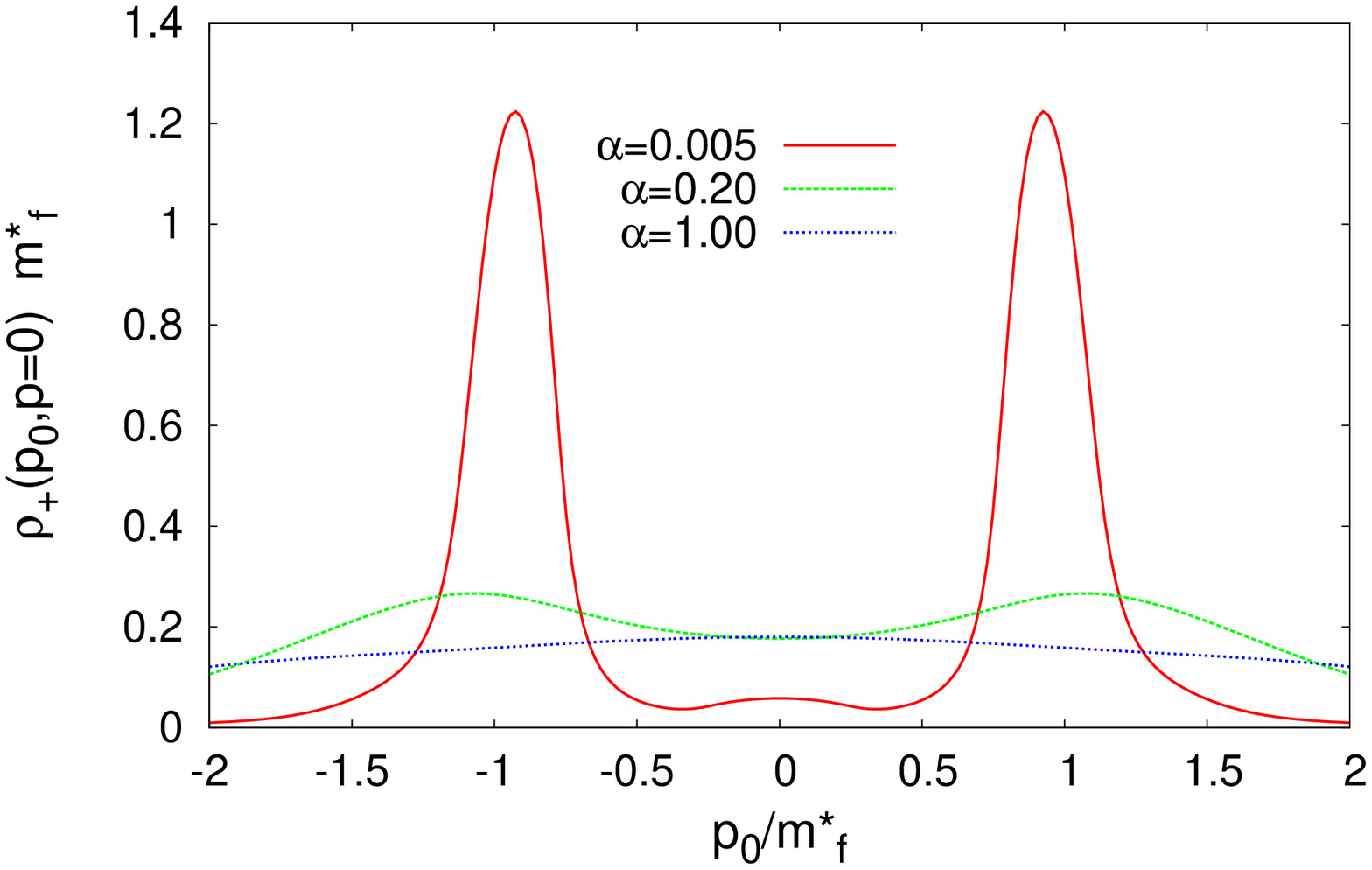}}
  \caption{The spectral density at $p=0$, $\alpha_+(p_0, p=0)$, at $T=0.3$, 
in the weak coupling region $\alpha=0.005$, in the intermediate coupling 
region $\alpha=0.20$ and in the strong coupling region $\alpha=1.0$.}
\label{rho_mfs}
\end{figure}
\begin{figure}[htbp] 
  \centerline{\includegraphics[width=7.5cm]{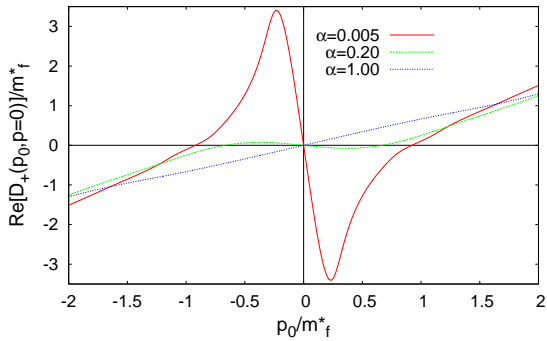}} 
  \caption{The real part of the inverse fermion propagator at $p=0$, Re$[D_+(p_0, p=0)]$, at $T=0.3$,
in the weak coupling region $\alpha=0.005$, in the intermediate coupling region $\alpha=0.20$ 
and in the strong coupling region $\alpha=1.0$.}
\label{dr_mfs}
\end{figure}
\begin{figure}[htbp] 
  \centerline{\includegraphics[width=7.5cm]{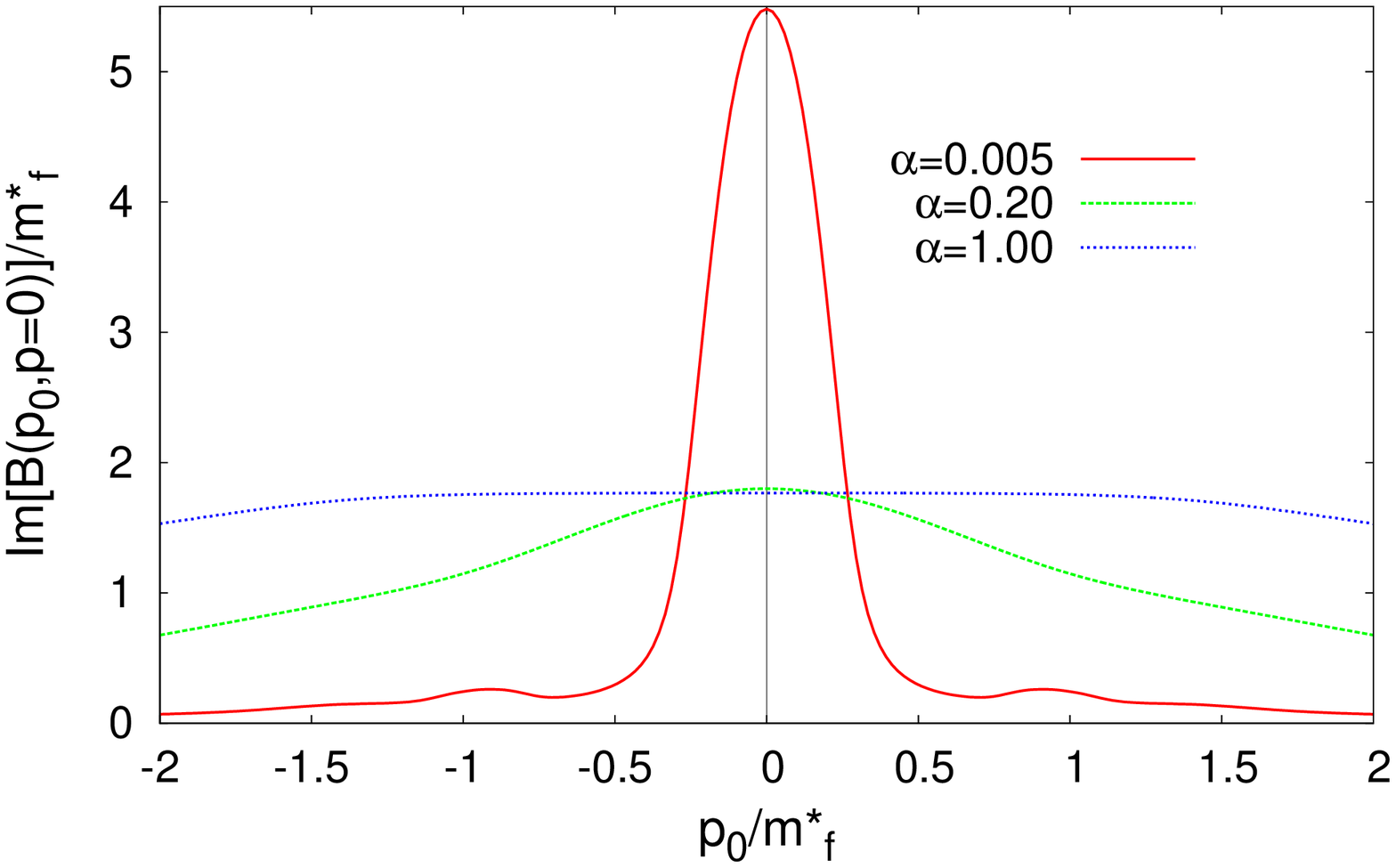}}
  \caption{The imaginary part of the inverse fermion propagator at $p=0$, Im$[D_+(p_0, p=0)]$=Im$[B (p_0, p=0)]$, 
at $T=0.3$, in the weak coupling region $\alpha=0.005$, in the intermediate coupling 
region $\alpha=0.20$ and in the strong coupling region $\alpha=1.0$.}
\label{di_mfs}
\end{figure}

Here let us see more carefully the structure of spectral density at $p=0$, $\rho_+(p_0, p=0)$, 
as a function of $p_0$. The $p_0$ coordinate of the peak position of $\rho_+(p_0, p=0)$ will give 
the mass of the corresponding mode. Figure~\ref{rho_mfs} shows the spectral densities, 
$\rho_+(p_0, p=0)$, at moderately high temperature $T=0.3$, in the weak coupling region 
$\alpha$\hspace{0.3em}\raisebox{0.4ex}{{$<$}\hspace{-0.75em}\raisebox{-.7ex}{$\sim$}}\hspace{0.3em}$0.01$, 
in the region of intermediate coupling strength 
$\alpha \approx 0.1 \sim 0.2$, and in the strong coupling region $\alpha \sim 1$. 

In order to see what actually happens during the transition from the triple peak structure 
in the weak coupling region to the double-peak one in the region of intermediate coupling 
strength, and finally to the single peak one at strong couplings, we present in 
Figs.~\ref{dr_mfs} and \ref{di_mfs} the Re[$D_+(p_0, p=0)$] and Im[$B(p_0, p=0)$] at $T=0.3$, respectively, 
both of which show three curves corresponding to the three regions of the coupling as in Fig.~\ref{rho_mfs}.

At weak coupling $\alpha=0.005$, we can clearly see in Fig.~\ref{rho_mfs} two sharp peaks at positive 
and negative $p_0$, representing the quasifermion and the plasmino poles. 
The center positions of both of peaks are in fact at $p_0/m^*_f \sim \pm 1$, which in fact almost 
coincide with the solution of the on-shell condition Re$[D_+(p_0, p=0)]=0$, as can be easily 
seen in Fig.~\ref{dr_mfs}. Thus at the weak coupling and high temperature both the 
quasifermion and the plasmino modes have a common thermal mass $m^*_f$, Eq.~\ref{eq_mf_star}, 
which is, as already noted, determined through the next-to-leading order calculation of HTL resummed 
effective perturbation theory~\cite{Rebhan,Schulz}. 

We can also barely recognize the existence of a slight ``peak'' around $p_0=0$, 
corresponding to a massless pole. The existence of this massless mode is also indicated by 
the on-shell condition Re$[D_+(p_0, p=0)]=0$, in which $p_0=0$ is always a solution. 
The problem with this third ``peak'' will be discussed in Sec.~\ref{sec:Solution-3rdPeak}.

It is also worth noticing that at weak coupling $\alpha=0.005$ Im$[B(p_0,p=0)]$ shows a
three-peak structure; a sharp steep peak centered at $p_0=0$, and two slight peaks centered
at $|p_0|=m^*_f$~\cite{footnote_peak}. All three peaks in Im$[B]$ appear corresponding
to the on-shell point Re$[D_+(p_0,p=0)]=0$, as was the case in the peaks of the 
spectral density, the sharpness of the peaks being completely turned over.

At intermediate coupling $\alpha=0.2$, the fermion spectral density exhibits a typical 
double-peak structure. The existence of these peaks, however, is not easy to understand. 
As we can easily make sure by comparing Figs.~\ref{rho_mfs} and \ref{dr_mfs}, they 
do not have exact correspondence to the poles 
of the fermion propagator, i.e., the zero point of the inverse propagator Re$[D_+(p_0, p=0)]=0$. 

At strong coupling $\alpha=1.0$, the situation becomes very simple. There exists only a broad 
single peak, whose existence is indicated by the on-shell condition Re$[D_+(p_0, p=0)]=0$, in 
which $p_0=0$ is the only solution in the strong coupling region (see Fig.~\ref{dr_mfs}). It is not clear 
whether the massless peak at strong couplings is exactly the same one at weak couplings noted 
above, or not. We will discuss this massless mode in Sec.~\ref{sec:Solution-3rdPeak}. 

To understand the typical structure of the fermion spectral density explained above, it is always 
important to correctly take notice of the height of the peak and the width of the corresponding pole, 
namely the height of the peak and the width of Im$[B(P)]$, given in Fig.~\ref{di_mfs}, in connection 
with the solution of the on-shell condition Re$[D_+(p_0, p=0)]=0$. 

In this sense the appearance of the double-peak structure at intermediate couplings is somewhat 
confusing. It is because, while there is a clear correspondence between the triple peaks at small 
couplings and the physical poles or modes (i.e., the quasifermion, the plasmino and the massless 
or ultrasoft modes), the double peaks at intermediate couplings do not have an obvious correspondence 
to the physical poles or modes.  They may correspond to the quasifermion and the plasmino modes, but the 
center positions of the peaks are apparently bigger than expected from the value of thermal mass, see, Fig.~\ref{rho_mfs}. 
In addition, as explained above, the third peak corresponding to the massless or the ultrasoft mode can not be recognized 
at all. This problem might arise from the broad-peak structure of the imaginary part of the mass function, Im$[B(p_0,p)]$, 
centered at $p_0=0$, see, Fig.~\ref{di_mfs}, and will be  discussed later in Secs.~\ref{sec:Solution-Peak},  
and \ref{sec:Solution-Dispersion} and Appendix~\ref{ap:Disp-Law-rho}.

With the appearance of this problem, it seems better to determine the position of the quasifermion pole by 
the solution of the on-shell condition Re$[D_+(p_0,p)]=0$ than by the peak position of the spectral density. 

\subsection{\label{sec:Solution-Peak}What determines the peak position of spectral density? Or the relation between 
the peak position of spectral density $\rho_+(P)$ and the zero point of the inverse propagator Re$[D_+(P)]$}

Here we study the problem, What determines the peak position of spectral density? At the end of the 
last section, \ref{sec:Solution-Density-Structure}, we briefly commented on the problem by focusing on the relation between 
the peak position of spectral density $\rho_+(P)$ and the zero point of the inverse propagator Re$[D_+(P)]$. 
There we also noticed that we should correctly take into account the information on the Im$[B(P)]$. 

Let us summarize what we have disclosed. a) At weak couplings there are two sharp peaks located 
at $p_0/m^*_f \sim \pm 1$, which in fact almost coincide with the solution of the on-shell 
condition Re$[D_+(p_0, p=0)]=0$. The third slight ``peak'' around $p_0 = 0$ is also indicated 
by the on-shell condition Re$[D_+(p_0, p=0)]=0$, in which $p_0=0$ is always a solution. 
b) Typical double peaks at intermediate strength of coupling do not have an obvious correspondence 
to the zero point of the inverse propagator Re$[D_+(p_0, p=0)]=0$. c) At strong couplings, 
there exists only one broad ``peak,'' whose existence is indicated by the on-shell 
condition Re$[D_+(p_0, p=0)]=0$, in which $p_0=0$ is the only solution in the strong coupling region. 

 There are two questions. (1) What happens at intermediate strength of coupling? 
(2) What causes the huge difference of the peak height between sharp peaks of fermion 
and plasmino modes and the slight peak of massless mode at weak couplings?  

We can add one more question: Does the massless peak (or the pole) at strong couplings 
represent the same massless mode at weak couplings? 
This third question, however, will be discussed in a separate paper. 

Now let us study questions (1) and (2) in order.

On question (1): First let us see the solution of the on-shell condition Re$[D_+(p_0,p=0)]=0$. 
In the range of intermediate couplings around $\alpha \sim 0.2$ (temperature is fixed at $T=0.3$), the real part 
of the chiral symmetric mass function at $p=0$, Re$[B(p_0,p=0)]$, as a function of $p_0$ exhibits 
a subtle structure around the origin. In studying the small $p_0$ region, it has a steep valley/peak 
structure at weak couplings, but as the coupling becomes stronger this valley/peak structure 
eventually diminishes in size and begins to behave almost as a straight line. 

The intermediate coupling region is the transition region: As the coupling gets stronger the two 
solutions of the on-shell condition
at $ |p_0|  \neq 0$ eventually approach $p_0=0$ and coincide with the solution 
at $p_0=0$ that always exists irrespective of the strength of the coupling. Thus the number of 
solutions of the on-shell condition Re$[D_+ (p_0,p=0)]=0$ changes suddenly from three to one. 

We should check here, in the considered region at $T=0.3$ with couplings around $\alpha=0.2$ and 
stronger, where the real part of the inverse propagator vanishes, i.e., where the solutions of 
Re$[D_+ (p_0,p=0)]=0$ exist. There are three solutions, two of them sit at $p_0 \neq 0$, i.e.,  
$|p_0| \simeq 0.6$ and the third one at $p_0=0$ (see Fig.~\ref{dr_mfs}), thus indicating the existence 
of three poles, or the appearance of three peaks in the spectral density. 

We should then see the 
shape and the position of the peak of Im$[B(p_0,p=0)]$. Im$[B(p_0,p=0)]$ always has a single broad 
peak around $p_0=0$ in the corresponding region of temperatures and couplings, i.e., $T=0.3$ and 
couplings around $\alpha=0.2$ and stronger (see Fig.~\ref{di_mfs}). 

With these facts we understand 
that, at intermediate coupling $\alpha=0.2$, the peak structure of Im$[B(p_0,p=0)]$ plays an 
important role, scratching out (washing away) the peak of the spectral density at $p_0=0$, and 
the not-so-steep but still Gaussian decreasing structure of Im$[B(p_0,p=0)]$ makes the positions 
of the  peaks of the spectral density at $|p_0| \simeq 0.6$ shift to larger $|p_0|$ values. 

On question (2): To understand this question, let us see Figs.~\ref{rho_mfs}, \ref{dr_mfs} 
and \ref{di_mfs} in the weak coupling region. 
The spectral density exhibits two sharp peaks at $p_0 \sim \pm m^*_f$, and one barely slight peak 
at $p_0=0$. The positions of these three peaks exactly agree with the three solutions of the 
on-shell condition Re$[D_+(p_0, p=0)]=0$ at $p_0 \sim \pm m^*_f$ and at $p_0=0$; thus these 
three modes rigidly correspond to the fermion, plasmino and massless modes, 
respectively~\cite{Klimov,Weldon,Kitazawa,Nakk-Nie-Pire}.  

In contrast with the spectral density, the structure of Im$[B(p_0,p=0)]$ (=Im$[D_+(p_0,p=0)])$ is simple. Im$[B(p_0,p=0)]$ at weak coupling $\alpha=0.005$, 
as can be seen in Fig.~\ref{di_mfs}, exhibits a sharp peak at $p_0=0$ and two slight peaks
at $|p_0|/m^*_f \simeq 1$. These peaks have a clear correspondence to the three solutions
of the on-shell condition Re$[D_+(p_0,p=0)]=0$. At the positions of two sharp peaks in 
the spectral density it is essential 
that Re$[D_+]$ is zero, and the imaginary part of it, or Im$[B]$, is so small that it 
does not play any essential 
role in the structure of the spectral density. At the position of the massless pole $p_0=0$,
however,  Im$[B]$ is so large that it plays an important role to 
almost scratch out the fact that Re$[D_+]$ is zero;
thus the peak structure almost disappears at around the origin.

The peak height at the pole is determined by its pole residue. This fact means that, by
 measuring the ratio of peak heights between the sharp peak representing the quasifermion 
mode and the slight peak at the origin representing the massless or ultrasoft mode, we can 
determine the ratio between the corresponding pole residues. 
This analysis will be carried out in a separate paper. 

\subsection{\label{sec:Solution-Dispersion}The quasifermion pole and quasifermion 
dispersion law}

\subsubsection{\label{sec:Solution-Dispersion-define}How to define the quasifermion pole}

Generally speaking, the pole of the propagator or the point where the inverse propagator
vanishes defines the corresponding particle and its dispersion law.  In the case of the thermal 
quasiparticle, however, its mass term usually has a finite, not small but on most occasions
quite large imaginary part, namely, the pole position of the thermal quasiparticle sits deeply inside 
the complex $p_0$ plane.
 
Because it is not very simple to study the structure of such a pole sitting deeply inside the
complex $p_0$ plane, we usually study such a pole by defining the condition so that the real
part of the inverse propagator vanishes as the on-shell condition. We adopt this definition of on-shell
throughout this analysis; then the quasifermion pole is defined by the zero point of the
real part of the chiral invariant fermion inverse propagator Re[$D_{\pm} (P) \equiv D_{\pm} (p_0, p)$],
\begin{equation}
   \mbox{Re}[D_{\pm} (p_0, p)] = 0  \ \ \ \ \mbox{at} \ \ \ \  (p_0 = \omega_{\pm}, p) ,
\label{eq_def_omega_by_ReD}
\end{equation}
which determines the dispersion law of this pole, $\omega_{\pm}(p)$.

There is of course another definition of on-shell and its corresponding pole. One  such
definition is to use the peak position of spectral density as the pole position of the
corresponding particle. With this definition we can also determine the dispersion law of this
pole, $\omega^{\rho}_{\pm} (p)$.  Though in most cases these two definitions give the 
same results, i.e., the dispersion law determined through the on-shell condition 
Re[$D_{\pm} (p_0 = \omega_{\pm}, p)] = 0$ agrees with the one determined by the peak
position of the spectral density, in some cases two definitions give different results.
We have also discussed in Sec.~\ref{sec:Solution-Peak} above, the possibility that
the peak position of the spectral density in the region of intermediate coupling strength
may not correctly represent the physical modes. We will discuss this problem in 
Appendix~\ref{ap:Disp-Law-rho}. Therefore, as mentioned above, we adopt 
Eq.~(\ref{eq_def_omega_by_ReD}) as the definition of on-shell.

\subsubsection{\label{sec:Solution-Dispersion-weak}Quasifermion dispersion law in the
weakly coupled QCD/QED medium and the fermion thermal mass}

\begin{figure}[htbp] 
  \centerline{\includegraphics[width=7.5cm]{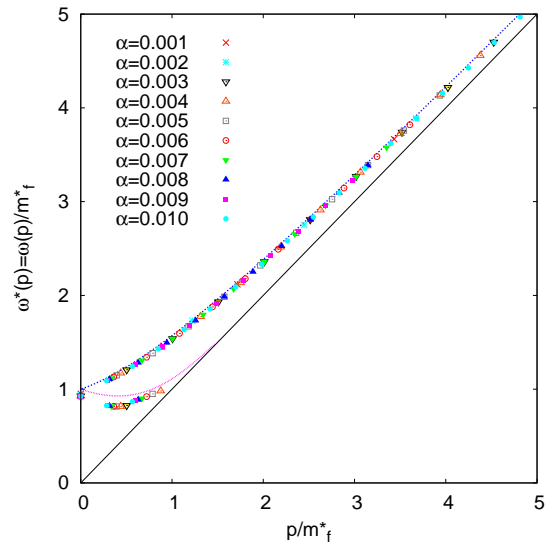}} 
  \caption{The normalized quasifermion dispersion law $\omega^*(p) \equiv \omega(p)/m^*_f$ at $T=0.3$ 
in the small coupling region.}
\label{dis_zeros_weak}
\end{figure}

Now let us study the quasifermion dispersion law determined through the on-shell condition,
Eq.~(\ref{eq_def_omega_by_ReD}), i.e., Re$[D_{\pm} (p_0 = \omega_{\pm}, p)] = 0$.
In Fig.~\ref{dis_zeros_weak} we give the quasifermion dispersion law $\omega =\omega_{\pm}(p)$
at small coupling and  at moderately high temperature.
It should be noted that, as can be seen in Fig.~\ref{dis_zeros_weak}, in the region
of weak coupling strength $\alpha  \hspace{0.3em}\raisebox{0.4ex}{$<$}\hspace{-0.75em}\raisebox{-.7ex}{$\sim$}\hspace{0.3em} 0.01$ 
the dispersion law lies on a universal curve determined by
the HTL calculations~\cite{Klimov,Weldon}.  Thus the result shows a good agreement with the HTL
resummed effective perturbation calculation. 

The important point is that both the quasifermion
energy $\omega_+ (p)$ and the plasmino energy $\omega_-(p)$ approach the same fixed value
$m^*_f$, Eq.~(\ref{eq_mf_star}), as $p \to 0$, namely, in Fig.~\ref{dis_zeros_weak} the normalized
energy $\omega^*_{\pm}(p) \equiv \omega_{\pm}(p)/m^*_f$ approaches 1 as $p \to 0$.
This fact clearly shows that the quasifermion as well as the plasmino have a definite thermal
mass $m^*_f$ of $O(gT)$ determined through the next-to-leading order calculation of HTL resummed
effective perturbation theory~\cite{Rebhan,Schulz}.  We should also note that the collective plasmino
mode exhibits a minimum at $p \neq 0$ and vanishes rapidly on to the light cone as $p$ gets large.

 It is also worth noticing that at weak coupling and moderately high temperature, the dispersion law in 
the small-$p$ region determined by the zero point of $D_+$ agrees well and almost coincides with 
that determined by the peak of the spectral density $\rho_+$. This fact can be understood by 
the one already noted in Sec.\ref{sec:Solution-Density-Structure} that in the weak coupling region 
the thermal mass of quasifermion determined by the peak position of spectral density almost 
coincides with the one determined by the solution of the on-shell condition Re$[D_+(p_0, p=0)]=0$, 
and that the thermal mass thus determined is $m^*_f$. The sharp peak structure of Im$[B(p_0, p=0)]$, 
or the narrow width structure of the quasifermion pole in the corresponding region may guarantee 
this fact. As the momentum $p$ becomes large, however, a discrepancy appears between them, 
especially in the dispersion law of the plsmino branch (see Appendix~\ref{ap:Disp-Law-rho}).

\subsubsection{\label{sec:Solution-Dispersion-Vanishing}Vanishing of the thermal mass in
the strongly coupled QCD/QED medium}

Next let us study how the result shown in Fig.~\ref{dis_zeros_weak} changes as the
coupling gets stronger, namely, in the region of intermediate to strong couplings.
For this purpose, let us see carefully the fermion branch of the quasifermion dispersion law in the small
momentum region. (N.B. Temperatures and couplings we are studying belong to the chiral
symmetric phase.)

\begin{figure}[htbp] 
  \centerline{\includegraphics[width=7.5cm]{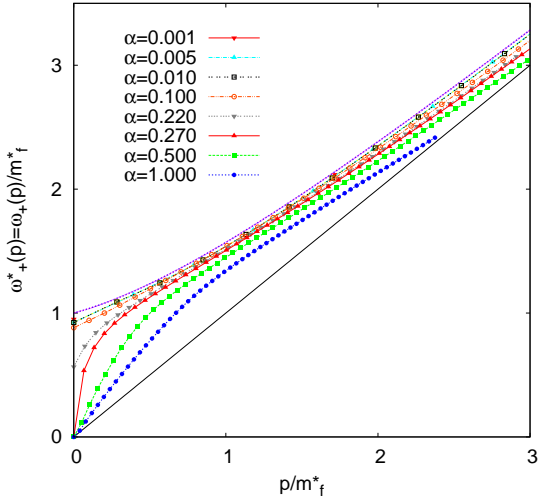}} 
  \caption{The$\alpha$ dependence of the normalized quasifermion dispersion 
law $\omega^*(p)\equiv \omega(p)/m^*_f$ at $T=0.3$, as the coupling becomes stronger. }
\label{dis_zeros_strong}
\end{figure}

Figure~\ref{dis_zeros_strong} shows the $\alpha$ dependence of the normalized 
dispersion law at $T=0.3$ as the coupling $\alpha$ becomes stronger, where the normalization
scale is the next-to-leading order thermal mass $m^*_f$.  In Fig.~\ref{dis_zeros_strong} 
we can clearly see, though in the weak coupling region we get the solution in good agreement with
the HTL resummed perturbation analyses, as the coupling becomes stronger from the intermediate to
strong coupling region the normalized thermal mass $\omega^*_+(p=0) \equiv \omega_+(p=0)/m^*_f$
begins to decrease from 1 and finally tends to zero
($\alpha  \hspace{0.3em}\raisebox{0.4ex}{$>$}\hspace{-0.75em}\raisebox{-.7ex}{$\sim$}\hspace{0.3em} 0.27$
in Fig.~\ref{dis_zeros_strong}).
Namely, in the thermal QCD/QED medium, the thermal mass of the quasifermion begins to decrease as the strength
of coupling gets stronger and finally disappears in the strong coupling region.
This fact strongly suggests that in the recently produced strongly coupled QGP the thermal mass of the quasifermion 
should vanish or at least become significantly lighter compared to the value in the ideal weakly coupled QGP.

\begin{figure}[htbp] 
  \centerline{\includegraphics[width=7.5cm]{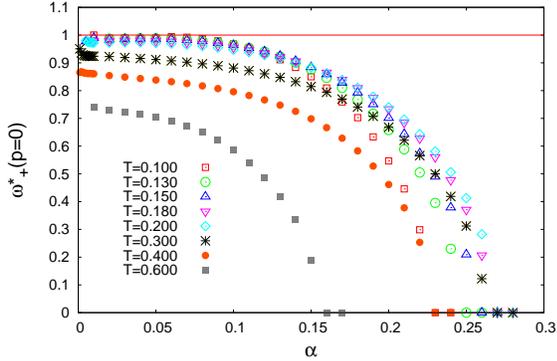}} 
  \caption{The $\alpha$ dependence of the normalized thermal mass $\omega^*_+(p=0)$. (See text.)}
\label{dis_zeros_a_tfix_mfs}
\end{figure}

To see the above behavior of the thermal mass more clearly, in Fig.~\ref{dis_zeros_a_tfix_mfs} we
show the normalized mass $\omega^*_+ (p=0)$ as a function of $\alpha$.  In the small coupling
region ($\alpha  \hspace{0.3em}\raisebox{0.4ex}{$<$}\hspace{-0.75em}\raisebox{-.7ex}{$\sim$}\hspace{0.3em} 0.1$)
 and around the temperature $T=0.1 \sim 0.2$, results of the thermal
mass agree well with those of the HTL resummed perturbation calculation.  As the coupling
gets stronger from intermediate to strong coupling regions, however, the normalized thermal
mass $\omega^*_+ (p=0)$ begins to decrease from 1 and finally goes down to zero; i.e.,  the
thermal mass vanishes.
\begin{figure}[htbp] 
  \centerline{\includegraphics[width=7.5cm]{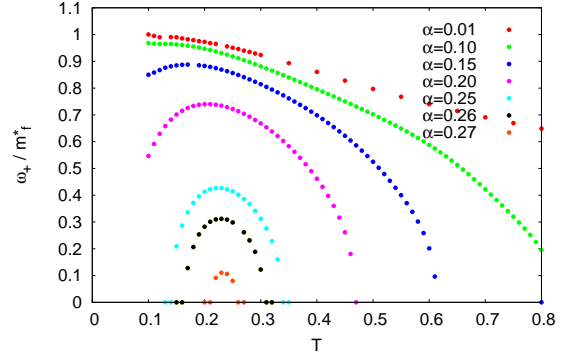}} 
  \caption{The $T$ dependence of the normalized thermal mass $\omega^*_+(p=0)$. (See text.)}
\label{dis_zeros_t_afix_mfs}
\end{figure}

Analogous behavior of thermal mass $\omega_+(p = 0)$ appears in the temperature dependence. As can be
seen in Fig.~\ref{dis_zeros_t_afix_mfs}, with any coupling $\alpha$, thermal mass decreases from $m^*_f$ as 
the temperature becomes higher, and finally at extreme high temperature $\omega_+(p = 0)$ becomes zero; 
thus the thermal mass vanishes. Figure~\ref{dis_zeros_t_afix_mfs} shows another
characteristic behavior as $T  \to$ small.  Almost at any coupling the thermal mass 
decreases and finally tends to vanish as temperature becomes lower.  This behavior is
consistent with the fact that at zero- temperature the thermal mass must vanish.  The
unexpected behavior is that, as the coupling becomes stronger, the thermal mass 
$\omega_+(p=0)$ vanishes at low but nonzero finite temperature.

Here it is to be noted that the ratio $\omega_+(p = 0)/T$ is not necessarily 
a constant and the $T$ dependence at not-so-high $T$ observed in our analysis is a consequence of the 
nonperturbative DSE analysis. It is because the additional dimensionful parameter, such as the regularization 
(or the cutoff) scale or the renormalization scale comes into the theory through the regularization and/or 
the renormalization of massless thermal QCD/QED. In our case the cutoff scale $\Lambda$ is introduced 
into the theory. The thermal mass, in fact, has a logarithmic $T$ dependence in the effective perturbation 
calculation; see, e.g., Rebhan's lecture in Ref.~\cite{Rebhan}.

\begin{figure}[htbp] 
  \centerline{\includegraphics[width=7.5cm]{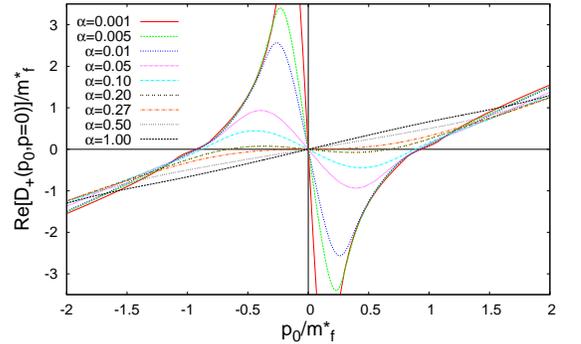}} 
  \caption{The $\alpha$ dependence of the real part of the inverse fermion propagator 
at $p=0$, Re$[D_+(p_0, p=0)]$=Re$[p_0 + B (p_0, p=0)$], at $T=0.3$.}
\label{ReD_peq0}
\end{figure}

The behavior of the thermal mass is determined by the behavior of the chiral invariant mass function
Re$[B(p_0, p)]$.  In Fig.~\ref{ReD_peq0} we show, for the sake of convenience, the 
$\alpha$ dependence of Re$[D_+ (p_0, p=0)]=$Re$[p_0+B(p_0, p=0)]$ at $T=0.3$.  At small
coupling Re$[D_+(p_0, p=0)]$ has a steep valley/peak structure in the small $p_0$ region,
but as the coupling becomes stronger this structure eventually disappears and 
Re$[D_+ (p_0, p=0)]$ belongs to behavior almost as a straight line with a slope $+1$.

Thermal mass is given by the solution of Re$[D_+(p_0, p=0)]=0$, i.e., the $p_0$ coordinate of the
intersection point of the drawn curve of Re$[D_+(p_0, p=0)]$ and the $p_0$ axis.
At first we can see with this figure that at small couplings there are three intersection
points, the one with positive $p_0$, the one with negative $p_0$, and the one at the
origin $p_0=0$, which correspond to the quasifermion, the plasmino and the massless (or 
ultrasoft) modes~\cite{Kitazawa,NYY11}, respectively.  

As the coupling becomes stronger  
($\alpha  \hspace{0.3em}\raisebox{0.4ex}{$>$}\hspace{-0.75em}\raisebox{-.7ex}{$\sim$}\hspace{0.3em} 0.27$ at $T=0.3$),
however, the number of the intersection points suddenly reduces and there appears only one
intersection point at $p_0=0$, which may correspond to the massless pole in the fermion propagator.  Thus we can
understand the behavior in  Fig.~\ref{dis_zeros_a_tfix_mfs}; namely, in the weak coupling region
$\omega^*_+(p=0) \equiv \omega_+(p=0)/m^*_f$ is almost unity, and reduces to zero in the strong coupling region
 ($\alpha  \hspace{0.3em}\raisebox{0.4ex}{$>$}\hspace{-0.75em}\raisebox{-.7ex}{$\sim$}\hspace{0.3em} 0.27$ at $T=0.3$), 
showing that the fermion thermal mass vanishes completely in the corresponding strong coupling region.

\subsubsection{\label{sec:Solution-Dispersion-Plasmino}Disappearance of the plasmino mode in
strongly coupled QCD/QED medium}

Finally we study what happens in the plasmino mode in the strongly coupled QCD/QED medium, by explicitly
examining the plasmino branch of the dispersion law. In  
Sec.~\ref{sec:Solution-Dispersion-Vanishing}, where we see the thermal mass vanish in the
strong coupling region, we only studied the structure of
the fermion branch of the dispersion law, and of the inverse propagator  at $p=0$,
 Re$[D_+(p_0, p=0)]=0$.  We cannot exactly see what happens in the plasmino mode without 
explicitly studying the plasmino branch of the dispersion law.

\begin{figure}[htbp] 
  \centerline{\includegraphics[width=7.5cm]{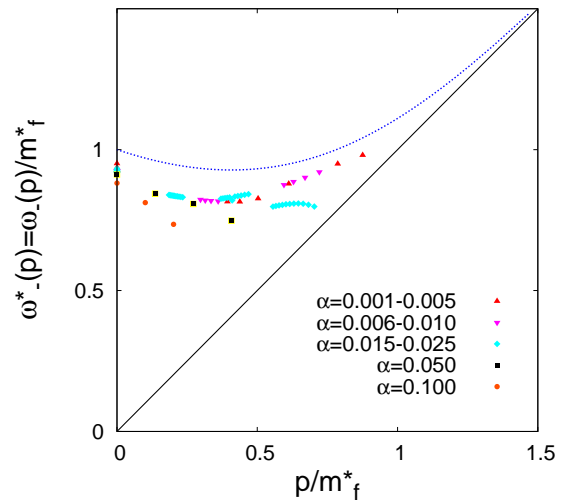}} 
  \caption{The $\alpha$ dependence of the normalized dispersion law $\omega_-^*(p) \equiv \omega_-(p)/m^*_f$  
of the plasmino branch at $T=0.3$.
The curve shown is the dispersion law of the plasmino branch in the HTL
calculation~\cite{Klimov}.}
\label{dis_plasmino}
\end{figure}

Figure~\ref{dis_plasmino} shows the $\alpha$ dependence of the normalized dispersion law
of the plasmino branch at $T=0.3$ as the coupling $\alpha$ becomes stronger. (At weak couplings
we already saw its structure in Fig.~\ref{dis_zeros_weak}.)

Paying attention to the plasmino branch, we recognize that, as the coupling gets stronger, the 
valley structure of the plasmino dispersion law, or the existence of the minimum in the plasmino
 dispersion law, observed in the weak coupling region, eventually disappears, and that the plasmino 
dispersion law sharply drops onto the light cone as the momentum $p$ becomes large. 

At $T=0.3$ in the small coupling region $\alpha
\hspace{0.3em}\raisebox{0.4ex}{$<$}\hspace{-0.75em}\raisebox{-.7ex}{$\sim$}\hspace{0.3em} 0.01$ the plasmino branch lies on the universal curve determined by the HTL calculation.
Around $\alpha \simeq 0.02$ the dispersion law of the plasmino branch begins to
change its structure: first the behavior as $p \to$ large begins to show sudden decrease
onto the light cone, then, second, around $\alpha \simeq 0.05$ the valley structure of
the plasmino dispersion law eventually disappears and the plasmino dispersion law 
monotonically drops sharply onto the light cone, and finally in the region $\alpha
\hspace{0.3em}\raisebox{0.4ex}{$>$}\hspace{-0.75em}\raisebox{-.7ex}{$\sim$}\hspace{0.3em} 0.27$ (at $T=0.3$) the plasmino branch totally disappears.

If the coupling gets further stronger, the thermal mass begins to decrease and eventually disappears 
at $\alpha \hspace{0.3em}\raisebox{0.4ex}{$>$}\hspace{-0.75em}\raisebox{-.7ex}{$\sim$}\hspace{0.3em} 0.27$, 
as noted in Sec.~\ref{sec:Solution-Dispersion-Vanishing}. The plasmino branch disappears at 
 $\alpha \hspace{0.3em}\raisebox{0.4ex}{$>$}\hspace{-0.75em}\raisebox{-.7ex}{$\sim$}\hspace{0.3em} 0.27$
also, which we can see in Fig.~\ref{dis_plasmino}, and the three modes, i.e., the fermion, the plasmino and the 
ultrasoft modes, finally merge and become a single massless mode that can be hardly detected as a real 
physical mode in the strongly coupled QGP, as noted before because of its large decay width.

\subsection{\label{sec:Solution-Thermal_mass}Thermal mass of the quasifermion}

In Sec.~\ref{sec:Solution-Dispersion} we have disclosed unexpected behavior of the thermal mass 
of the quasifermion in the strong coupling QCD/QED, namely the fact that the thermal mass vanishes in the 
strongly coupled QCD/QED medium (or, the recently produced strongly coupled QGP). Also we have pointed 
out that at weak coupling and high temperature both the quasifermion and the plasmino have a common thermal 
mass $m^*_f$, Eq.~(\ref{eq_mf_star}), determined through the next-to-leading order calculation of HTL resummed 
effective perturbation theory~\cite{Rebhan,Schulz}.

\begin{figure}[htbp] 
  \centerline{\includegraphics[width=7.5cm]{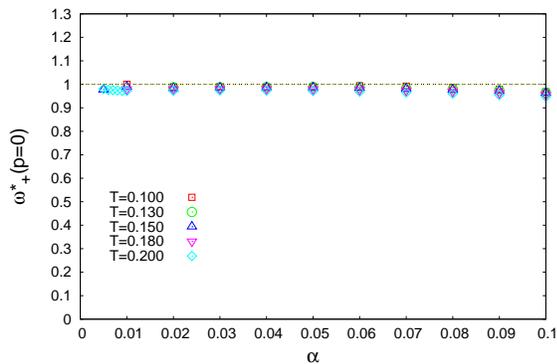}} 
  \caption{The $\alpha$ dependence of the normalized thermal mass $\omega^*_+(p=0)$. (See text.)}
\label{dis_zeros_a_tfix_mfs_rescale}
\end{figure}
In this section we examine how accurately the thermal mass $m^*_f$, Eq.~(\ref{eq_mf_star}), can describe the 
thermal mass calculated in our analysis. Figure~\ref{dis_zeros_a_tfix_mfs}, showing the coupling dependence of the 
thermal mass presented in the last section, covers a wide range of couplings and gives us only a rough image, 
and thus is not suited to the present purpose. 

Here we present Fig.~\ref{dis_zeros_a_tfix_mfs_rescale}, the rescaled 
version of Fig.~\ref{dis_zeros_a_tfix_mfs}, showing the thermal mass calculated in our analysis in the weak coupling region
 $\alpha \hspace{0.3em}\raisebox{0.4ex}{$<$}\hspace{-0.75em}\raisebox{-.7ex}{$\sim$}\hspace{0.3em} 0.1$.
Now we can see clearly that in the whole region of the temperature
 $0.100  \hspace{0.3em}\raisebox{0.4ex}{$<$}\hspace{-0.75em}\raisebox{-.7ex}{$\sim$}\hspace{0.3em}  T
 \hspace{0.3em}\raisebox{0.4ex}{$<$}\hspace{-0.75em}\raisebox{-.7ex}{$\sim$}\hspace{0.3em} 0.200$
at weak couplings 
 $\alpha \hspace{0.3em}\raisebox{0.4ex}{$<$}\hspace{-0.75em}\raisebox{-.7ex}{$\sim$}\hspace{0.3em} 0.1$,
the normalized thermal mass $\omega^*_+(p=0) \equiv \omega_+(p=0)/m^*_f$ is almost unity, namely, the
thermal mass calculated in our analysis, $\omega_+(p = 0)$, is well described by $m^*_f$. 
As the temperature becomes higher, discrepancy becomes evident and larger; 
the normalized thermal mass $\omega^*_+(p=0)$ deviates from unity and gets smaller, --namely, 
$\omega_+(p = 0)$ begins to decrease from $m^*_f$ and becomes smaller.

It should be noted, however, while at very small couplings 
 $\alpha \hspace{0.3em}\raisebox{0.4ex}{$<$}\hspace{-0.75em}\raisebox{-.7ex}{$\sim$}\hspace{0.3em} 0.01$
a common tendency can be recognized in Fig.~\ref{dis_zeros_a_tfix_mfs}
that the normalized thermal mass $\omega^*_+(p=0)$ approaches unity, 
except at extreme high temperatures. 

In studying the temperature dependence of the thermal mass, which is shown in Fig.~\ref{dis_zeros_t_afix_mfs}, another fact can be 
recognized. The first thing that attracts our attention is that, except in the small coupling region, the normalized 
thermal mass $\omega^*_+(p = 0)$ shows a peak structure, namely, that $\omega^*_+(p = 0)$ decreases as  
the temperature both becomes higher and becomes lower. This fact, in the former case we already noted 
in Sec.~\ref{sec:Solution-Dispersion-Vanishing},
is unexpected and not easy to understand with the knowledge we have learned through the effective perturbation
analyses. The behavior in the lower temperature region may indicate that the thermal mass shows a behavior 
proportional to $T / \log(1/T)$, while in the high temperature region the
thermal mass shows a behavior proportional to $T \log(1/T)$. (N.B.: The temperature $T$ 
varies in the range $0 \le T \le 1$.) 

As for the temperature dependence in the higher temperature region, we can only say 
definitely at present that the 
ratio $\omega_+(p = 0)/T$ is not necessarily a constant and the $T$ dependence of the normalized thermal 
mass $\omega^*_+(p = 0)$ observed in our analysis is a consequence of the nonperturbative DSE analysis. 

At small couplings the normalized thermal mass $\omega^*_+(p = 0)$ seems to approach unity as the temperature 
becomes lower, showing the well-known behavior of the thermal mass being proportional to the temperature $T$, 
and is easy to understand.

\subsection{\label{sec:Solution-3rdPeak}Existence of the third peak, or the ultrasoft mode}

The quasifermion and the plasmino modes are well understood in the HTL resummed 
analyses, the latter being the collective mode to appear in the thermal environment. 
What is the third peak? Is it nothing but convincing evidence of the existence of a massless 
or an ultrasoft mode? Is there any signature in our analysis?

The existence of the massless or the ultrasoft fermionic mode has been suggested first in the
one-loop calculation~\cite{Kitazawa} when a fermion is coupled with a massive boson with
mass $m$. The spectral function of the fermion gets to have a massless peak in addition to
the normal fermion and the plasmino peaks. Recently a possible existence of collective
fermionic excitation in the ultrasoft energy-momentum region
 $p \hspace{0.3em}\raisebox{0.4ex}{$<$}\hspace{-0.75em}\raisebox{-.7ex}{$\sim$}\hspace{0.3em} 
g^2T$ has been investigated analytically through
perturbative calculation~\cite{Kitazawa,Hidaka}. 
Both of these analyses are confined to the
weak coupling regime, and nothing is known about what happens in the sQGP we are interested in. In this
sense first we will study the structure of the third mode, i.e., of the massless or the ultrasoft mode,
in the weakly coupled QCD/QED medium, and then we will proceed to the intermediate and strong coupling region to 
investigate how the ultrasoft mode behaves in such an environment~\cite{NYY11}.

\begin{figure}[htbp] 
  \vspace*{-0.5cm}
  \centerline{\includegraphics[width=7.5cm]{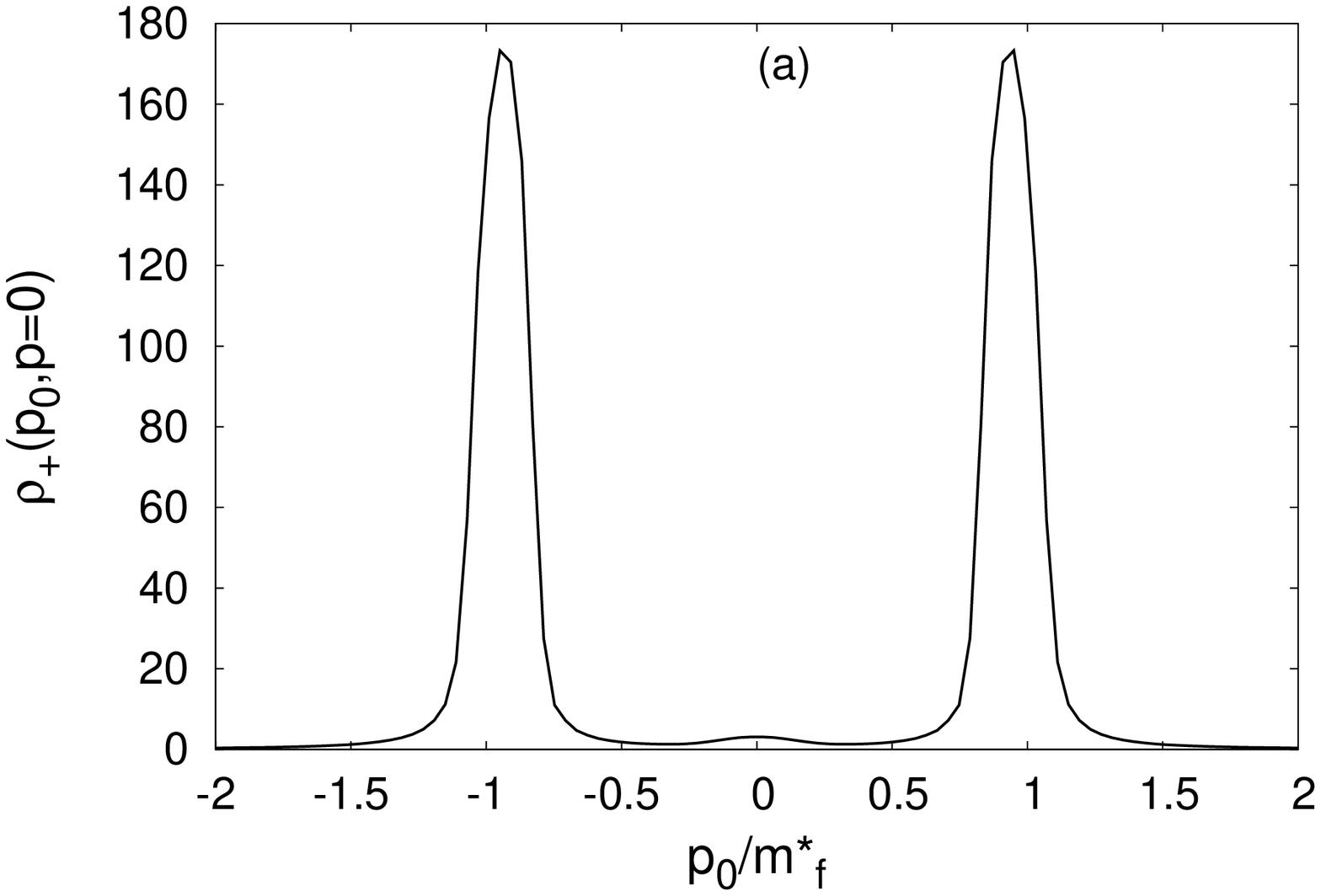}} 
  \centerline{\includegraphics[width=7.5cm]{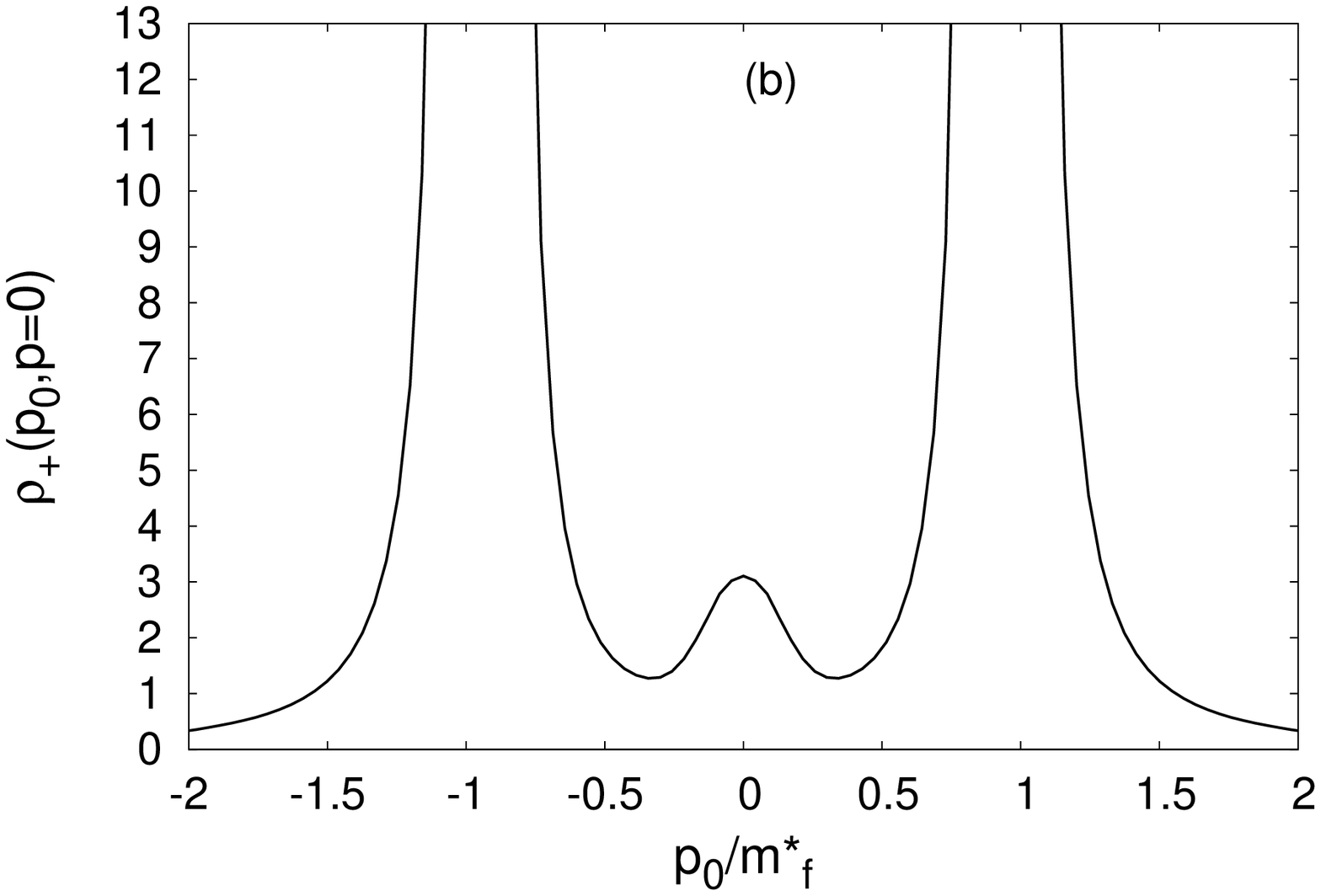}}
\caption{(a) Quasifermion spectral density $\rho_+(p_0,p=0)$ at small coupling region 
($\alpha=0.001, T=0.3$). \ (b) Quasifermion spectral density $\rho_+(p_0,p=0)$ enlarged 
around the origin.}
\label{rho_t030a0001}
\end{figure}

Now let us study the structure of the third mode, i.e., of the massless or the ultrasoft mode,
in the weakly coupled QCD/QED medium. First we give in Fig.~\ref{rho_t030a0001}(a) the structure of
spectral density $\rho_+ (p_0, p=0)$ at the weak coupling region ($\alpha=0.001, T=0.3$).
Two sharp peaks, representing the quasifermion and the plasmino poles are clearly seen,
and the existence of a slight ``peak'' can also be recognized around $p_0=0$.
To see more clearly, Fig.~\ref{rho_t030a0001}(b) shows a rescaled version of  Fig.~\ref{rho_t030a0001}(a), 
where we can clearly see the ``peak'' 
structure around $p_0=0$. This third peak is nothing but convincing evidence of the existence
of a massless or an ultrasoft mode~\cite{Kitazawa,NYY11}. This peak is indistinctively slight
compared to the sharp quasifermion and plasmino peaks.

Here we should take notice of the fact that the peak height (or more rigorously the 
integral of the peak over a finite peak width) of the ultrasoft mode
centered at $p_0=0$ is, roughly speaking,  $O(g)$ lower than the peak height of the normal fermion or
the plasmino peak centered at $p_0=\pm m^*_f$.
This rough result does not exactly agree with  
what Hidaka \textit{et al.} have shown in their works~\cite{Hidaka} concerning the 
residue of the ultrasoft fermion mode, and we will perform a more detailed analysis on this problem
in a separate paper.

\subsection{\label{sec:Solution-DecayWidth}Decay width of the quasifermion, or the imaginary part 
of the chiral invariant mass function $B$}

Finally let us study the decay width of the quasifermion, or  
the imaginary part of the chiral invariant mass function Im[$B(p_0, p)$] at $p=0$. 
The decay width of the quasifermion is extensively studied through the HTL resummed effective 
perturbation calculation~\cite{Nakk-Nie-Pire}, giving a gauge-invariant result of $O(g^2T \log (1/g))$. 
However, as is shown above, the quasiparticle exhibits an unexpected behavior, such as the 
vanishing of the thermal mass in the strongly coupled QCD/QED medium, completely different from 
that expected from the HTL resummed effective perturbation analyses. How does the decay width 
of the quasifermion exhibit its property in the corresponding strongly coupled QCD/QED medium?

\begin{figure}[htbp] 
  \centerline{\includegraphics[width=7.5cm]{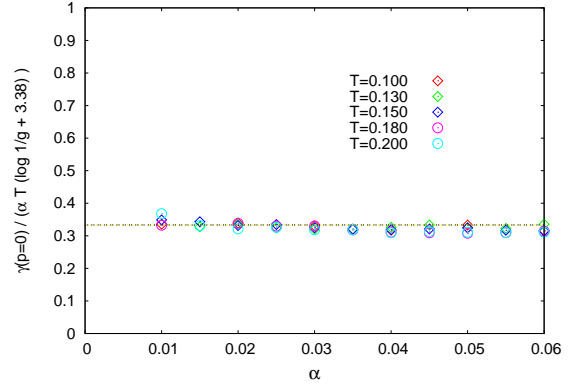}}
  \caption{The decay width of the quasifermion at rest $\gamma(p=0)$ in the weakly coupled QGP. 
The dotted straight line Eq.~(\ref{eq_gamma_approx}) represents the result from the HTL 
resummed calculation. (See text.)}
  \label{Decay-width-small}
\end{figure}
\begin{figure}[htbp] 
  \centerline{\includegraphics[width=7.5cm]{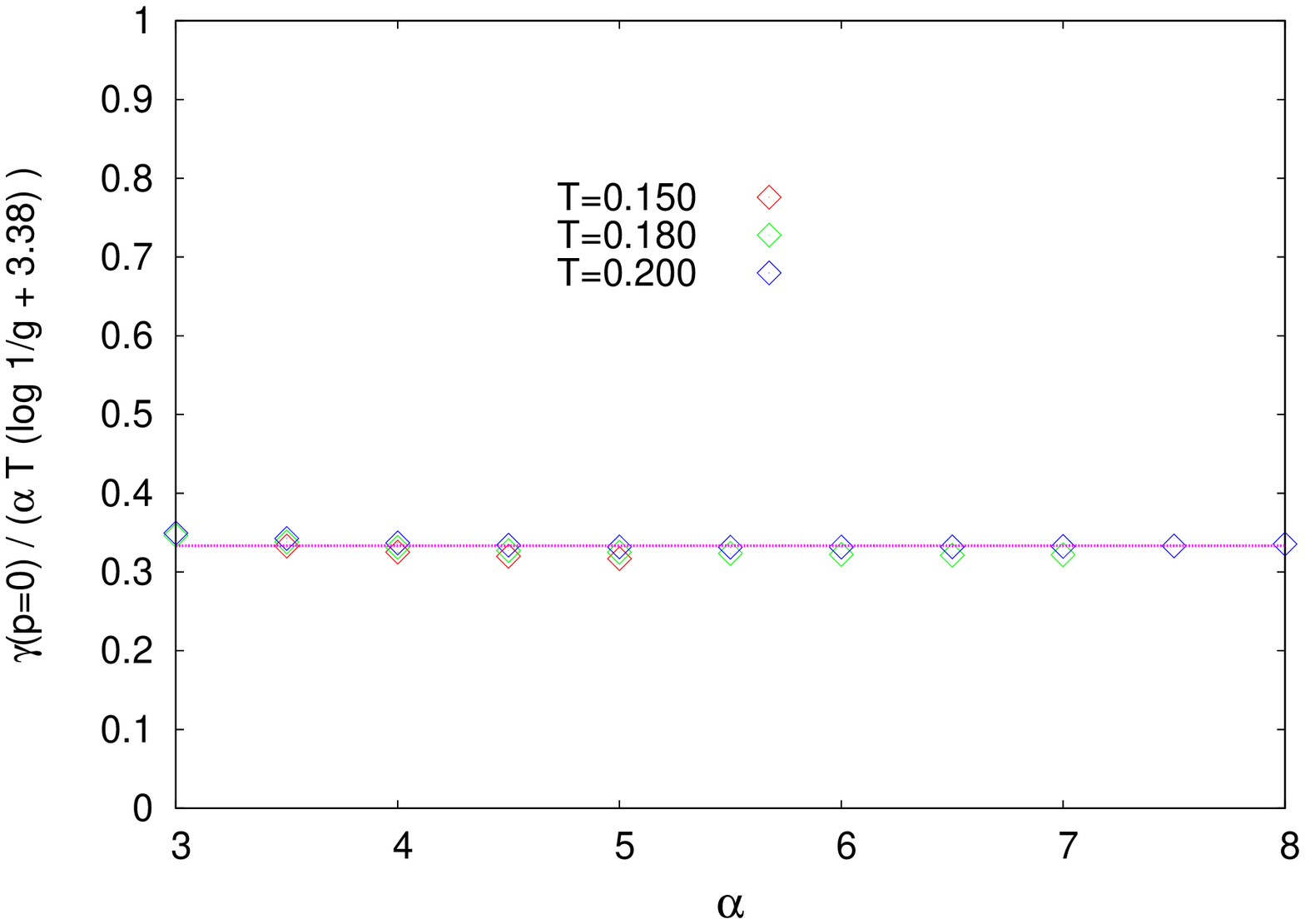}}
  \caption{The decay width of the quasifermion at rest $\gamma(p=0)$ in the strongly coupled QGP. 
The horizontal straight line Eq.~(\ref{eq_gamma_approx}) represents the result from the HTL resummed calculation. 
(See text.)}
  \label{Decay-width-large}
\end{figure}

In Figs.~\ref{Decay-width-small} and \ref{Decay-width-large} we show the decay width
of the quasifermion $\gamma (p)$ at $p=0$, in the weakly coupled and in the strongly
coupled QGP, respectively, where
\begin{eqnarray}
\label{eq_gamma_def}
\gamma (p) &\equiv& \frac12 \mbox{Im}[D_+(p_0=\omega_+(p), p)] \nonumber \\
& & \times \left[ \left .\frac{\partial}{\partial p_0} \mbox{Re}[D_+(p_0, p)] 
\right|_{p_0=\omega_+(p)} \right]^{-1} .
\end{eqnarray}
In both figures, the fitting straight line represents
\begin{eqnarray}
\label{eq_gamma_approx}
\gamma (p=0) &=& \frac13 \alpha T \left( \log \frac{1}{g} + c \right), \\
c &\simeq& 3.38.  \nonumber
\end{eqnarray}

In the weak coupling and high temperature QGP, the decay width $\gamma (p=0)$,
Eq.~(\ref{eq_gamma_approx}), agrees with the HTL resummed 
effective perturbation calculation~\cite{Nakk-Nie-Pire} up to a numerical 
factor~\cite{footnote-Decay-width} (see Fig.~\ref{Decay-width-small}).

Quite unexpectedly even in the strongly coupled QGP, the resulting decay width
$\gamma (p=0)$, Eq.~(\ref{eq_gamma_approx}), shows the totally same behavior of 
$O(g^2T \log(1/g))$ as in the weakly coupled QGP up to the numerical factor
and the $O(g^2T)$ correction term, Fig.~\ref{Decay-width-large}.

\begin{figure}[htbp] 
\centerline{\includegraphics[width=7.5cm]{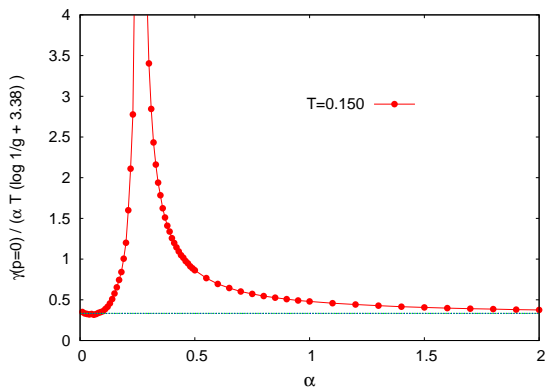}} 
\caption{The $\alpha$ dependence of the decay width of the quasifermion at 
rest $\gamma(p=0)$ at $T=0.15$.}
\label{Decay-width-intermediate}
\end{figure}

What happens in the intermediate coupling region? The results of the decay width at
$T=0.150$ are given in Fig.~\ref{Decay-width-intermediate}. From this figure we can understand
how the decay width in the weak coupled QGP and the one in the strongly coupled QGP
coincide.  In the intermediate coupling region, the decay width of the quasifermion shows
a ``rich'' structure. The decay width $\gamma (p=0)$ diverges at the vanishing point
of the thermal mass $\omega_+(p=0)$, namely, the point where $\omega_+(p=0)$ 
first hits zero as the coupling changes [see Eq.~(\ref{eq_gamma_def})].

This behavior is again not expected, because the quasifermion in the small coupling and high 
temperature QGP and the one in the strong coupling and high temperature QGP are totally 
different; in the former case the quasifermion has a thermal mass of $O(gT)$ and the plasmino 
branch exists in a fermion dispersion law, while in the latter case thermal mass of the 
quasifermion vanishes and the plasmino branch disappears.

The temperature dependence of the decay width is again described by 
Eq.~(\ref{eq_gamma_approx}), namely, the decay width of the quasifermion is linearly
proportional to the temperature, both in the weak and strong coupling QGP.

\begin{figure}[htbp] 
  \centerline{\includegraphics[width=7.5cm]{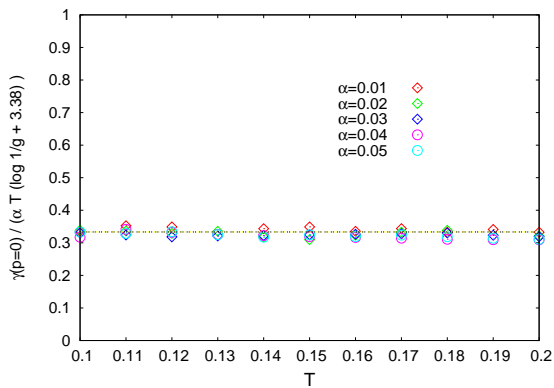}} 
  \caption{The $T$ dependence of the decay width of the quasifermion at rest 
$\gamma(p=0)$ in the weakly coupled QGP. The dotted straight line, Eq.~(\ref{eq_gamma_approx}), represents the result 
from the HTL resummed calculation. (See text.)}
  \label{Decay-width-small-afix}
\end{figure}

\begin{figure}[htbp] 
  \centerline{\includegraphics[width=7.5cm]{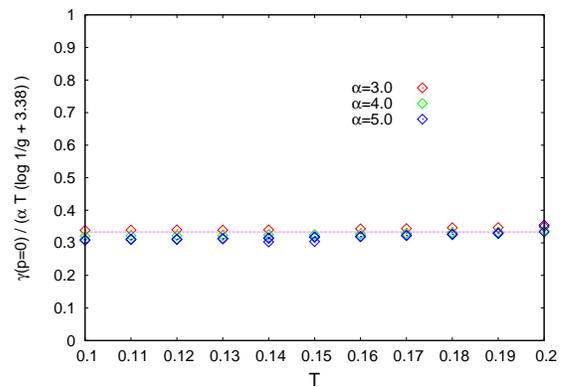}}
  \caption{The $T$ dependence of the decay width of the quasifermion at rest $\gamma(p=0)$ 
in the strongly coupled QGP. The horizontal straight line, Eq.~(\ref{eq_gamma_approx}), represents 
the result from the HTL resummed calculation. (See text.)}
  \label{Decay-width-large-afix}
\end{figure}

This behavior can be clearly 
seen in Figs.~\ref{Decay-width-small-afix} and \ref{Decay-width-large-afix}, and also in Figs.~\ref{Decay-width-small} and~\ref{Decay-width-large}. 
The former fact indicates that in the strongly coupled QGP, recently produced at RHIC and LHC, 
the predicted massless or the ultrasoft pole is very hard to be detected as a real physical mode.


\section{\label{sec:Discussion}Summary and Discussion}
In this paper we carried out a nonperturbative analysis of a thermal quasifermion in 
thermal QCD/QED by studying its self-energy function through the DSE
 with the HTL resummed improved ladder kernel. With the solution of 
the DSE we studied the properties of the thermal quasifermion spectral density and 
its peak structure, as well as the dispersion law of the physical modes corresponding to the 
poles of the thermal quasifermion propagator. Through the study of the quasifermion we 
elucidated the properties of thermal mass and the decay width of fermion and 
plasmino modes, and also paid attention to properties of the possible third mode, 
both especially in the strongly coupled QCD/QED medium.

What we have revealed in this paper is the drastic change of properties of 
the ``quasifermion'' depending on the strength of the interaction among constituents 
of the QCD/QED medium:

\begin{description}
\item{i)} \ \ In the weak coupling region, or in the weakly coupled QCD/QED medium:
 $\alpha \hspace{0.3em}\raisebox{0.4ex}{$<$}\hspace{-0.75em}\raisebox{-.7ex}{$\sim$}\hspace{0.3em} 0.02$ or
 $g \hspace{0.3em}\raisebox{0.4ex}{$<$}\hspace{-0.75em}\raisebox{-.7ex}{$\sim$}\hspace{0.3em} 0.5$  at $T=0.3$.
The on-shell conditions through the peak structure of spectral density and from 
the zero point of the quasifermion inverse propagator give the same structure 
and properties of the quasifermion. A rigid quasiparticle picture holds with 
the thermal mass $m^*_f$, Eq.~(\ref{eq_mf_star}), and a small imaginary part 
or the decay rate $\gamma \simeq  g^2T \log(1/g)$ and the fermion act as a 
basic degree of freedom of the medium. 
The thermal mass $m^*_f$ is nothing but the next-to-leading order result 
of the HTL resummed effective perturbation calculations~\cite{Rebhan}. 
A fermion and the plasmino mode appear. Thus the results in the weak 
coupling region well reproduce those of the HTL resummed effective 
perturbation calculations.

   The triple peak structure of the quasifermion spectral density is clearly 
observed, indicating the existence of the fermionic ultrasoft third mode, 
which is absent from the HTL resummed effective perturbation analyses.

\item{ii)} \ \ In the strong coupling region, or in the strongly coupled QCD/QED medium: 
 $\alpha \hspace{0.3em}\raisebox{0.4ex}{$>$}\hspace{-0.75em}\raisebox{-.7ex}{$\sim$}\hspace{0.3em} 0.1$ or
 $g \hspace{0.3em}\raisebox{0.4ex}{$>$}\hspace{-0.75em}\raisebox{-.7ex}{$\sim$}\hspace{0.3em} 1$ at $T=0.3$.
Both the spectral density and the inverse fermion propagator tell the single 
massless peak structure with large imaginary part, or, the decay rate
$\gamma \simeq g^2T \log(1/g)$. The quasiparticle picture \textit{\'a la} Landau has 
been broken down in the strongly coupled QCD/QED medium. The thermal mass 
vanishes and there appears only the fermion mode, the plasmino mode 
disappears in the strongly coupled medium.

\item{iii)} \ \ In the intermediate coupling region:
 $0.02 \hspace{0.3em}\raisebox{0.4ex}{$<$}\hspace{-0.75em}\raisebox{-.7ex}{$\sim$}\hspace{0.3em} 
   \alpha \hspace{0.3em}\raisebox{0.4ex}{$<$}\hspace{-0.75em}\raisebox{-.7ex}{$\sim$}\hspace{0.3em} 
   0.1$ or 
 $0.5 \hspace{0.3em}\raisebox{0.4ex}{$<$}\hspace{-0.75em}\raisebox{-.7ex}{$\sim$}\hspace{0.3em} g
  \hspace{0.3em}\raisebox{0.4ex}{$<$}\hspace{-0.75em}\raisebox{-.7ex}{$\sim$}\hspace{0.3em} 1$
  at $T=0.3$.
In this region the spectral density and the inverse fermion propagator 
tell a completely different structure. The spectral density tells that there 
should be two particle modes with large decay rates, while the inverse 
fermion propagator tells that there should be three poles in the propagator; 
thus there may exist three modes in this coupling region, just as in the case 
in the weakly coupled medium. We conclude that the indication of the inverse 
fermion propagator tells the truth, see the text. Anyway the intermediate 
coupling region is the transitional region for the fermion in the medium 
to behave as a rigid quasiperticle, acting as a basic degree of freedom in the medium.
\end{description}

Here we give several comments and discussion on the results of the present analysis.

(1) It is not \textit{a priori} very clear which one really defines the  
the physical on-shell particle and its dispersion law, the peak position of the 
quasifermion spectral density $\rho_{\pm}(p_0, p)$ or the zero point of the 
real part of the inverse fermion propagator Re$[D_{\pm}(p_0, p)]=0$, especially 
when the imaginary part is not very small. If we adopt the peak position of 
the quasifermion spectral density $\rho_{\pm}(p_0, p)$ as the on-shell point 
of the particle, then the corresponding dispersion law exhibits a branch 
developing into the spacelike domain of space-time. There is also the problem of 
the double peak structure of the quasifermion spectral density in the transitional 
intermediate coupling region, as noted above in (iii). With these facts we adopt 
Re$[D_{\pm}(p_0, p)]=0$ as the on-shell condition of the physical particle to 
study its dispersion law and various properties, such as the thermal mass 
and the decay width, etc. 

(2) With the on-shell condition Re$[D_{\pm}(p_0, p)]=0$ we select the particle 
mode and study its dispersion law and the particle properties. The on-shell 
condition at $p=0$, Re$[D_{\pm}(p_0, p=0)]=0$, always has  solution at $p_0=0$, 
which may correspond to the ultrasoft mode. This correspondence is, however, 
not so simple. The structure of the imaginary part around the on-shell point of 
the propagator plays an important role to make this correspondence exact. 
This relationship was pointed out by Kitazawa \textit{et al.}~\cite{Kitazawa}; 
if in the medium the bosonic mode with nonzero mass (with small decay rate) 
couples with the fermion, then the quasifermion spectral density shows a 
triple peak structure corresponding to the ultrasoft third mode together 
with the fermion and the plasmino modes. The appearance of the peak at $p_0=0$ 
corresponding to the ultrasoft third mode is guaranteed with the vanishment 
of the imaginary part, Im$[D_{\pm}(p_0, p=0)]=0$, at $p_0=0$, which happens 
because of the coupling of the fermion with the massive bosonic mode. Such 
a mechanism may not work in the QED medium since no massive bosonic 
excitations in the QED medium are expected.

In the present DSE analysis, the imaginary part of the fermion inverse propagator does not 
vanish at $p_0=0$, Im$[D_{\pm}(p_0, p=0)] \neq 0$, but rather shows a peak 
structure at $p_0=0$. This peak structure actually suppresses the peak height 
of the spectral density, as noted in the text, 
Sec.~\ref{sec:Solution-3rdPeak}. In this sense the appearance of the 
ultrasoft third peak with very low peak height in the present analysis 
may have a different origin from that of Kitazawa \textit{et al.}~\cite{Kitazawa} and 
from Hidaka \textit{et al.}~\cite{Hidaka}. This problem will be discussed further in a 
separate paper.

(3) We have noted that the thermal mass of the quasifermion decreases as the 
coupling gets stronger, and finally vanishes in the strong coupling region. 
This fact indicates that the thermal mass of the quasiquark vanishes and behaves 
as a massless fermion with a large decay rate in the recently discovered strongly
coupled QGP. It is not so simple, however, whether such a particle can be 
experimentally observed as a massless quasifermion or not.

(4) As noted above, the decay rate $\gamma(p)$ of the quasifermion in the QCD/QED 
medium shows a typical $g^2T \log(1/g)$ behavior both in the weakly and the 
strongly coupled medium. In the transitional intermediate coupling environment, 
however, the decay rate shows a rich structure and $\gamma(p=0)$ even diverges 
at the vanishing point of the thermal mass. It would be quite exciting if we could 
find some methods to be able to verify experimentally the unexpected behavior 
of the thermal mass and the decay rate.

\appendix
\section{\label{ap:HTL-DSE}Approximations to get the HTL resummed improved ladder DSE}

In the present analysis, we solve the DSE for the retarded fermion self-energy 
function $\Sigma_R$, with the HTL resummed  gauge boson propagator Eq.~(\ref{eq_DSE}), 
by adopting further the following two approximations to get Eq.~(\ref{eq_DSE_AB}): (i) the 
point-vertex approximation and (ii) the modified instantaneous exchange approximation
to get the final DSEs we solve, on which we give brief explanations below.

(i) Point-vertex approximation to get Eq.~(\ref{eq_DSE_AB}).

As for the vertex function $^*\Gamma^{\mu}$ we adopt the point-vertex approximation, namely we simply 
set $ ^*\Gamma^{\mu} = \gamma^{\mu}$ disregarding the HTL corrections to  $ ^*\Gamma^{\mu}$.
Thus we investigate the ladder (point-vertex) DS equation with the HTL resummed gauge boson propagator.

There are two reasons. First, without the point-vertex approximation the numerical calculation we should carry 
out becomes so complicated that we cannot manage with the power of the computer we use, because the 
HTL resummed contribution to the vertex function is the nonlocal interaction term, and also because it behaves 
singular in numerical calculations. Second, in the DSE with the HTL resummed vertex function, it is difficult 
to resolve the problem of double counting of diagrams~\cite{Aurenche}, especially at the level of numerical analyses. Being 
free from this problem in the numerical analysis is the main reason why we make use of the point-vertex approximation.

(ii) Modified instantaneous exchange approximation to get the final DSEs to solve.

The next approximation we make use of is the modified instantaneous exchange (IE) approximation (i.e., set the 
energy component of the gauge boson to be zero) to the gauge boson propagator  $^*G^{\mu \nu}$. 
The retarded ($R \equiv RA$) and correlation ($C \equiv RR$) components of the HTL resummed gauge 
boson propagator $^*G^{\mu \nu}$ are given by~\cite{Weldon} 
\begin{widetext}
\begin{subequations}
\label{eq_Gauge_Prop}
\begin{eqnarray}
\label{eq_Gauge_Prop_R}
{}^*G_R^{\mu\nu}(K) \!\! &\equiv& \!\! {}^*G_{RA}^{\mu\nu} (-K,K) \ =\ \frac{1}{{}^*\Pi^R_T(K) -K^2 - i \epsilon k_0} A^{\mu \nu}
    + \frac{1}{{}^*\Pi^R_L(K) -K^2 - i \epsilon k_0} B^{\mu \nu}
    - \frac{\xi}{K^2 + i \epsilon k_0} D^{\mu \nu} , \\
\label{eq_Gauge_Prop_C}
{}^*G_C^{\mu\nu} (K) &\equiv& {}^*G_{RR}^{\mu\nu} (-K,K) \ = \ 
    (1+2n_B(k_0)) \left[ {}^*G_R^{\mu \nu}(K) - {}^*G_A^{\mu \nu}(K) \right] , 
\end{eqnarray}
\end{subequations}
\end{widetext}
with $^*\Pi^R_T$ and $^*\Pi^R_L$ being the HTL contributions to the transverse and longitudinal modes of the retarded 
gauge boson self-energy, respectively~\cite{Klimov}. The parameter $\xi$ is the gauge-fixing parameter
 ($\xi=0$ in the Landau gauge). 

In the above, $A^{\mu \nu}$, $B^{\mu \nu}$ and $D^{\mu \nu}$ are the projection tensors given by~\cite{Weldon}
\begin{subequations}
\label{Tensor}
\begin{eqnarray}
\label{Tensor_A}
A^{\mu\nu} &=& g^{\mu\nu}-B^{\mu\nu} -D^{\mu\nu},  \\ 
\label{Tensor_B}
B^{\mu \nu} &=& -\frac{\tilde{K}^{\mu} \tilde{K}^{\nu}}{K^2},  \\
\label{Tensor_C} 
D^{\mu\nu} &=& \frac{K^{\mu} K^{\nu}}{K^2},  
\end{eqnarray}
\end{subequations}
where $\tilde{K}=(k,k_0 \hat{\bf k})$, $k=\sqrt{{\bf k}^2}$ and $\hat{\bf k}={\bf k}/k$ denote the unit three vector along ${\bf k}$.

The modified IE approximation we make use of  consists of taking the IE limit in the HTL resummed longitudinal (electric) gauge boson 
propagator, $^*G^{\mu \nu}_L$, that is proportional to $B^{\mu \nu}$, while keeping the exact HTL resummed form for the 
transverse (magnetic) gauge boson propagator, $^*G^{\mu \nu}_T$, that is proportional to $A^{\mu \nu}$, and also for the 
massless gauge term in proportion to $D^{\mu \nu}$. The reason why we do not take the IE limit to the transverse mode is 
that the IE approximation reduces the transverse mode to the pure massless propagation, and thus makes the important thermal 
effect, i.e., the dynamical screening of transverse propagation disappear.

With the above two approximations, we obtain the final HTL resummed improved ladder DSEs for the invarinat scalar 
functions $A$, $B$, and $C$, to solve.

\section{\label{ap:cutoff-dep}Cutoff dependence}

In this appendix we explain the cutoff dependence of the present analysis. 

As explained in Sec.~\ref{sec:DSE-Sigma}, in solving the DSEs, Eq.~(\ref{eq_DSE_AB}), 
we are forced to introduce a momentum cutoff in the integration over the four-momentum 
$\int d^4K$; $K=(k_0, \text{\bf k})$ is the fermion four-momentum.
The cutoff method we make use of is as follows ($\Lambda$ denotes an arbitrary cutoff parameter 
and plays a role to scale any dimensionful quantity, e.g., $T = 0.3$ means $T = 0.3\Lambda$):
\begin{center}
\begin{tabular}{llcl}
three-momentum & ~$\text{\bf k}$~ & ~:~ & $k=|\text{\bf k}| \le \Lambda$  \\
energy         & ~$k_0$~ & ~:~ & $|k_0| \le  \Lambda_0$ 
\end{tabular}
\end{center}
In the present analysis we make the ratio $r \equiv \Lambda_0/\Lambda$ vary $r=1 \sim 5$, 
and fix it so as to get a stable solution to the fermion spectral density. 

\begin{figure*}[htbp]
\begin{tabular}{ccc}
\begin{minipage}{0.47\hsize}
  \centerline{\includegraphics[width=8cm]{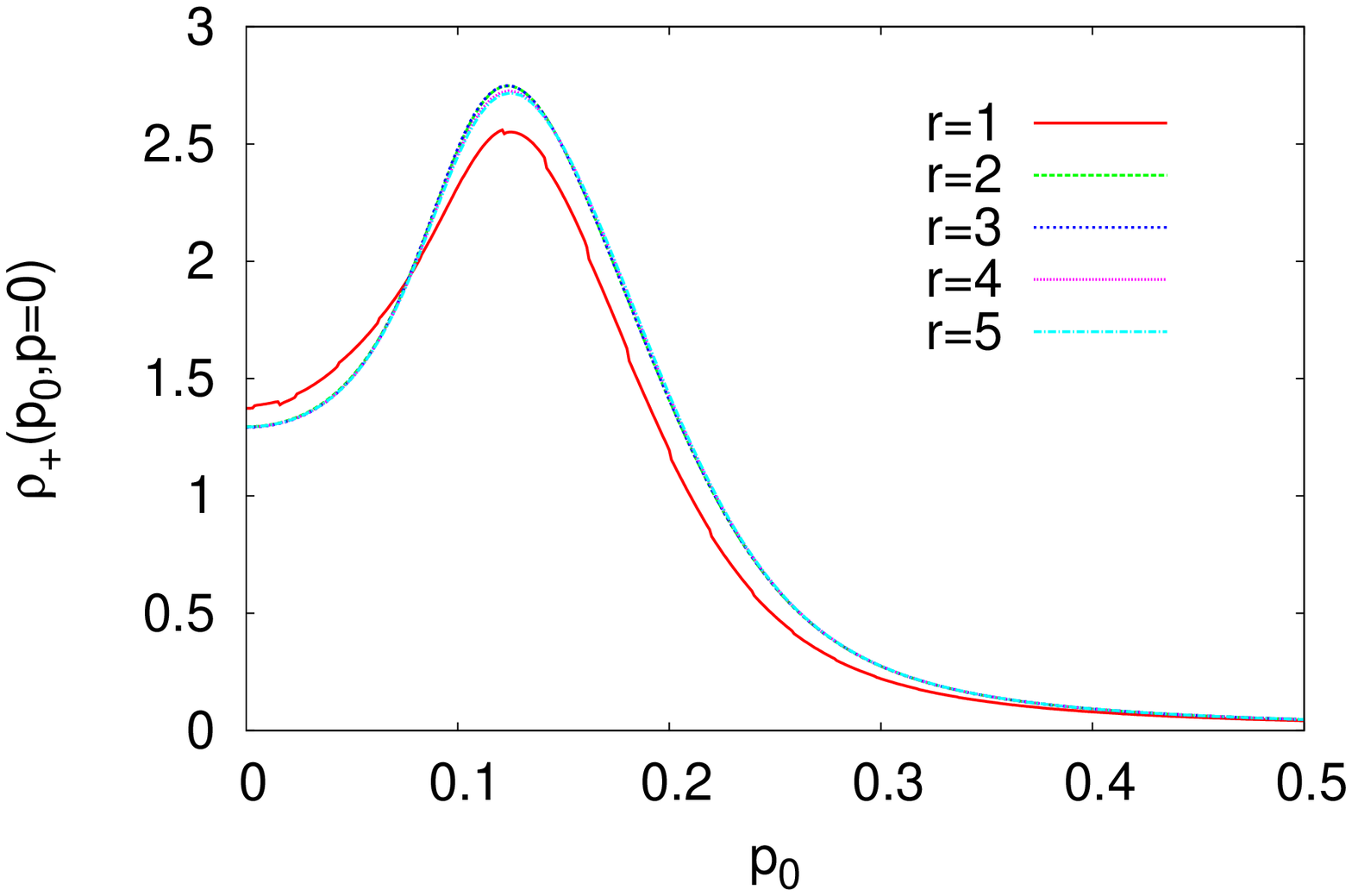}} 
  \caption{The cutoff dependence of the quasifermion spectral density at $p=0$, $\rho_+(p_0,p=0)$, 
at $T=0.4$ and $\alpha=0.1$. $r=\Lambda_0/\Lambda$.}
\label{cutoff_dep1}
\end{minipage}
\begin{minipage}{0.06\hsize}
\ \ \ 
\end{minipage}
\begin{minipage}{0.47\hsize}
  \centerline{\includegraphics[width=8cm]{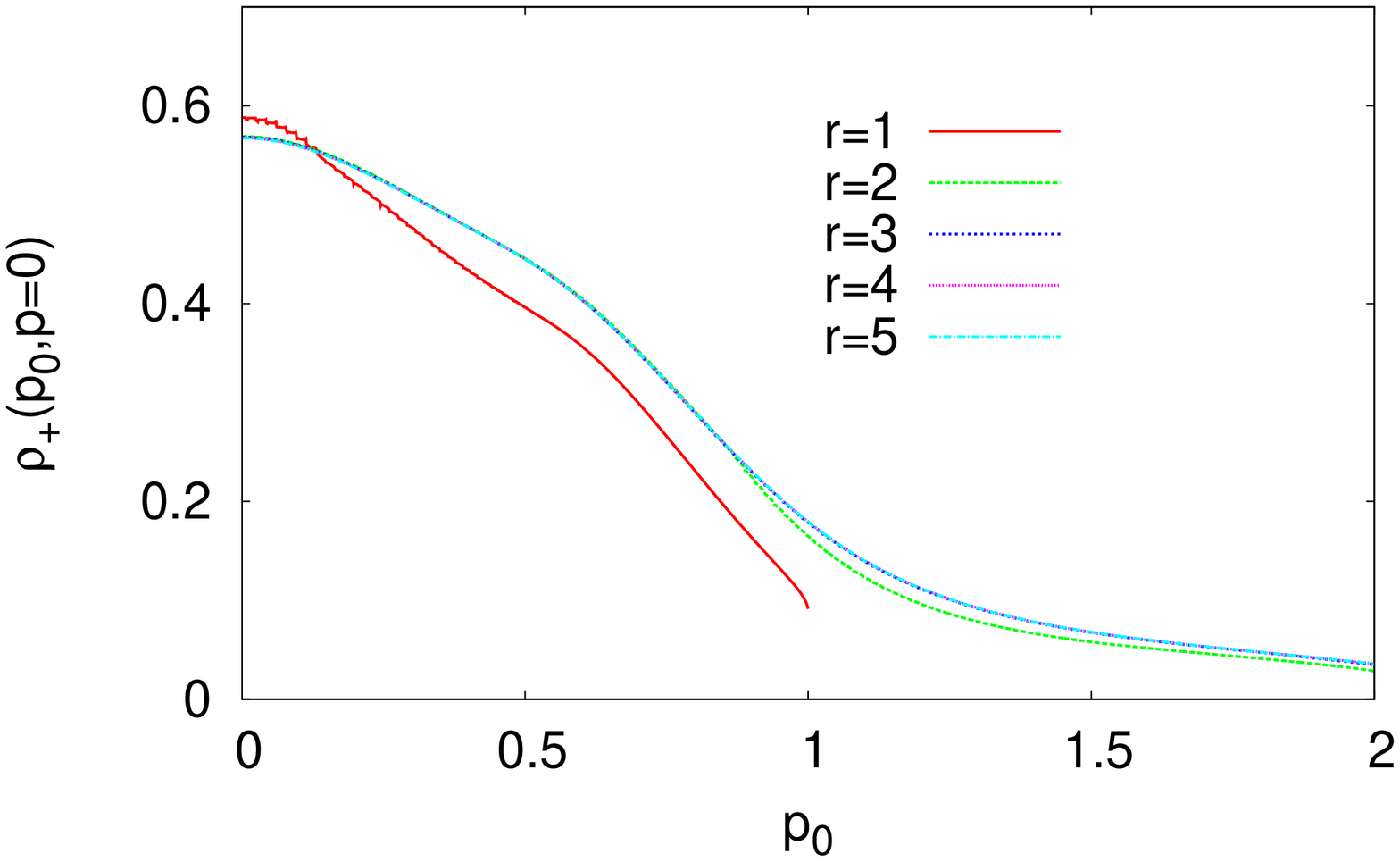}} 
  \caption{The cutoff dependence of the quasifermion spectral density at $p=0$, $\rho_+(p_0,p=0)$, 
at $T=0.4$ and $\alpha=1.0$. $r=\Lambda_0/\Lambda$.}
\label{cutoff_dep2}
\end{minipage}
\end{tabular}
\end{figure*}

In Fig.~\ref{cutoff_dep1} we show how the spectral density changes as a function of $k_0$ as we vary the ratio in the 
range $r=1 \sim 5$. We can easily recognize that at $T=0.4$ and $\alpha=0.1$ we can get a stable solution 
if we choose $r \ge 2$. 

The situation is almost the same but slightly differs at different
 $T$ and $\alpha$; see Fig.~\ref{cutoff_dep2}
at $T=0.4$, $\alpha=1.0$ and compare with Fig.~\ref{cutoff_dep1} at  $T=0.4$, $\alpha=0.1$.
As the coupling becomes stronger it is safe to choose larger values of $r$.

The stability of the solution can be checked by the saturation of the sum rules, Eqs.~(\ref{eq_sum_rule_1}) and (\ref{eq_sum_rule_2}). 
As already noted, the sum rule Eq.~(\ref{eq_sum_rule_3}) heavily relied on the HTL calculation; 
thus we do not use this sum rule. The result is given in Table~\ref{table-cutoff-dep}, again showing 
the stability of the solution \textit{when we 
choose $r \ge 2$} (or more safely $r \ge 3$).

With the above results we choose, in most cases except at very strong couplings,
an appropriate value of the ratio in the range $r \ge 2$, 
depending on the region of temperatures and/or couplings we study. 
The extreme high temperature may cause another problem, namely, the problem of simulation artifact; therefore we 
restrict the temperature to the region 
$T \hspace{0.3em}\raisebox{0.4ex}{$<$}\hspace{-0.75em}\raisebox{-.7ex}{$\sim$}\hspace{0.3em} 0.6$ 
and do not perform our analysis in an extreme high temperature region 
$T \hspace{0.3em}\raisebox{0.4ex}{$>$}\hspace{-0.75em}\raisebox{-.7ex}{$\sim$}\hspace{0.3em} 0.8$.
\begin{widetext}
\begin{center}
\begin{table}
\caption{The cutoff dependence of the saturation of the sum rules, Eqs.~(\ref{eq_sum_rule_1}) 
and (\ref{eq_sum_rule_2}): 
$r \equiv \Lambda_0/\Lambda$.\label{table-cutoff-dep}}
\begin{tabular}{|c|c|c|c|c|c|c|c|} \hline
\multicolumn{3}{|c|}{ } & $r=1$ & $r=2$ & $r=3$ & $r=4$ & $r=5$ \\ \hline
   & Eq.~(\ref{eq_sum_rule_1}) & 
       $p=0$ & 0.949 & 0.994 & 0.994  & 0.996 & 1.003 \\ \cline{3-8}
 $T=0.2$ &   &   $p=0.02$ & 0.986 & 0.985 & 0.983 & 0.982 & 0.990 \\ \cline{2-8}
 $\alpha=0.01$ &  Eq.~(\ref{eq_sum_rule_2}) &
       $p=0.02$ & 0.0197 & 0.0207 &0.0207  & 0.0209 & 0.0202 \\ \cline{3-8}
  & &    $p=0.04$ & 0.0392 & 0.0404 & 0.0404 & 0.0407 & 0.0404 \\ \cline{3-8}
   & &   $p=0.1$ & 0.0990 & 0.1009 &  0.1009 & 0.1004 & 0.1013 \\\hline
   & Eq.~(\ref{eq_sum_rule_1}) & 
       $p=0$ & 0.927 & 1.006 & 1.006 & 1.006 & 1.006 \\ \cline{3-8}
 $T=0.4$ &   &   $p=0.02$ & 0.929 & 1.006 & 1.006 & 1.006 & 1.006 \\ \cline{2-8}
 $\alpha=0.1$ &  Eq.~(\ref{eq_sum_rule_2}) &
       $p=0.02$ & 0.0180 & 0.0206 & 0.0207 & 0.0207 & 0.0207 \\ \cline{3-8}
  & &    $p=0.04$ & 0.0354 & 0.0406 & 0.0408 & 0.0408 & 0.0408 \\ \cline{3-8}
   & &   $p=0.1$ & 0.0948 & 0.1011 & 0.1016 & 0.1016 & 0.1016 \\\hline
   & Eq.~(\ref{eq_sum_rule_1})
    &   $p=0$ & 0.759 & 0.974 & 1.018 & 1.021 & 1.021 \\ \cline{3-8}
 $T=0.4$  & &      $p=0.02$ & 0.761 & 0.975 & 1.017 & 1.021 & 1.021 \\ \cline{2-8}
 $\alpha=1.0$ &    Eq.~(\ref{eq_sum_rule_2}) &
       $p=0.02$ & 0.0108 & 0.0171 & 0.0199 & 0.0202 & 0.0203 \\ \cline{3-8}
  & &    $p=0.04$ & 0.0218 & 0.0350 & 0.0407 & 0.0413 & 0.0413 \\ \cline{3-8}
   & &   $p=0.1$ & 0.0546 & 0.0879 & 0.1024 & 0.1037 & 0.1038 \\\hline
\end{tabular}
\end{table}
\end{center}
\end{widetext}

\section{\label{ap:Phase}Phase boundary in the Landau gauge}

In order to study the phase transition and to determine the phase boundary 
of thermal QCD/QED, we should solve the DSE for the retarded fermion 
self-energy function $\Sigma_R$, Eq.~(\ref{eq_Sigma}). 
For the present purpose, however, we must study the $\Sigma_R$ that has a 
$c$-number scalar mass function $C(P)$,
\begin{equation}
\label{eq_Sigma2}
   \Sigma_R(P) = (1-A(P)) p_i \gamma^i - B(P) \gamma^0 + C(P).
\end{equation} 

The DSE in the Landau gauge to determine the three scalar
invariants $A(P)$, $B(P)$ and $C(P)$ becomes coupled integral equations as follows:
\begin{widetext}
\begin{subequations}
\label{eq_DSE_ABC}
\begin{eqnarray}
\label{eq_DSE_AC}
 p^2[1-A(P)] &=& g^2 \left. \int \frac{d^4K}{(2 \pi)^4}
       \right[ \{1+2n_B(p_0-k_0) \} \text{Im}[\ ^*G^{\rho \sigma}_R(P-K)]
       \times  \nonumber \\
  & & \Bigl[ \{ K_{\sigma}P_{\rho} + K_{\rho} P_{\sigma}
       - p_0 (K_{\sigma} g_{\rho 0} + K_{\rho} g_{\sigma 0} ) 
       - k_0 (P_{\sigma} g_{\rho 0} + P_{\rho} g_{\sigma 0} )
       + pkz g_{\sigma \rho} \nonumber \\
  & & + 2p_0k_0g_{\sigma 0}g_{\rho 0} \}\frac{A(K)}{[k_0+B(K)+i
       \epsilon]^2 - A(K)^2k^2-C(K)^2 }
       + \{ P_{\sigma} g_{\rho 0} + P_{\rho} g_{\sigma 0} \nonumber \\
  & & - 2p_0 g_{\sigma 0} g_{\rho 0} \}
       \frac{k_0+B(K)}{[k_0+B(K)+i \epsilon]^2 - A(K)^2k^2-C(K)^2
       } \Bigr] + \{1-2n_F(k_0) \}
       \times \nonumber \\ 
  & & \ ^*G^{\rho \sigma}_R(P-K) \text{Im} \Bigl[
       \{ K_{\sigma}P_{\rho}  + K_{\rho} P_{\sigma} - p_0 (K_{\sigma}
       g_{\rho 0} + K_{\rho} g_{\sigma 0} ) - k_0 (P_{\sigma}
       g_{\rho 0} + P_{\rho} g_{\sigma 0} ) \nonumber \\
  & & + pkz g_{\sigma \rho} + 2p_0k_0g_{\sigma 0}g_{\rho 0}\}
       \frac{A(K)}{[k_0+B(K)+i \epsilon]^2 - A(K)^2k^2-C(K)^2} 
       \nonumber \\
  & & \left. +  \{ P_{\sigma} g_{\rho 0} + P_{\rho} g_{\sigma 0}
       - 2p_0 g_{\sigma 0} g_{\rho 0} \}
       \frac{k_0+B(K)}{[k_0+B(K)+i \epsilon]^2 - A(K)^2k^2-C(K)^2 } \Bigr] \right] \ , \\
\label{eq_DSE_BC}
 B(P) &=& g^2 \left. \int \frac{d^4K}{(2 \pi)^4} \right[
        \{1+2n_B(p_0-k_0)\} \text{Im}[\ ^*G^{\rho \sigma}_R(P-K)] \times
         \nonumber \\
  & & \Bigl[ \{ K_{\sigma} g_{\rho 0} + K_{\rho} g_{\sigma 0}
       - 2k_0 g_{\sigma 0} g_{\rho 0} \}
       \frac{A(K)}{[k_0+B(K)+i \epsilon]^2 - A(K)^2k^2-C(K)^2} \nonumber \\
  & & + \{ 2g_{\rho 0} 2g_{\sigma 0} - g_{\sigma \rho} \} 
       \frac{k_0+B(K)}{[k_0+B(K)+i \epsilon]^2 - A(K)^2k^2-C(K)^2 }
       \Bigr] + \{1-2n_F(k_0) \} \times \nonumber \\ 
  & & \ ^*G^{\rho \sigma}_R(P-K) \text{Im} \Bigl[ \frac{A(K)}{[k_0+B(K)+i
       \epsilon]^2 - A(K)^2k^2-C(K)^2 } 
       \{ K_{\sigma} g_{\rho 0} + K_{\rho} g_{\sigma 0}  \nonumber \\
  & & \left. - 2k_0 g_{\sigma 0} g_{\rho 0} \} + \frac{k_0+B(K)}{[k_0+B(K)+
       i \epsilon]^2 - A(K)^2k^2-C(K)^2}
       \{ 2g_{\rho 0} 2g_{\sigma 0} - g_{\sigma \rho} \} \Bigr] 
       \right] \ , \\
\label{eq_DSE_CC}
 C(P) &=& \left. g^2 \int \frac{d^4K}{(2 \pi)^4}  g_{\rho \sigma} \right[
        \{1+2n_B(p_0-k_0)\} \text{Im}[\ ^*G^{\rho \sigma}_R(P-K)] \times
         \nonumber \\
  & & \frac{C(K)}{[k_0+B(K)+i \epsilon]^2 - A(K)^2k^2-C(K)^2} + \{1-2n_F(k_0) \}
         \ \times \nonumber \\ 
  & & \left. \ ^*G^{\rho \sigma}_R(P-K) \text{Im} \Bigl[ \frac{C(K)}{[k_0+B(K)+i
       \epsilon]^2 - A(K)^2k^2-C(K)^2 } \Bigr] \right] \ . 
\end{eqnarray}
\end{subequations}
\end{widetext}
The above DSEs, Eq.~(\ref{eq_DSE_ABC}), may have several solutions, and 
we choose the ``true'' solution by evaluating the effective potential $V[S_R]$ 
for the fermion propagator function $S_R$, then finding the lowest energy solution. 
The effective potential $V[S_R]$ we evaluate is given in  Sec.~\ref{sec:DSE-potential}, 
Eq.~(\ref{eq_potential}).

\begin{figure}[htbp] 
 \centerline{\includegraphics[width=8.0cm]{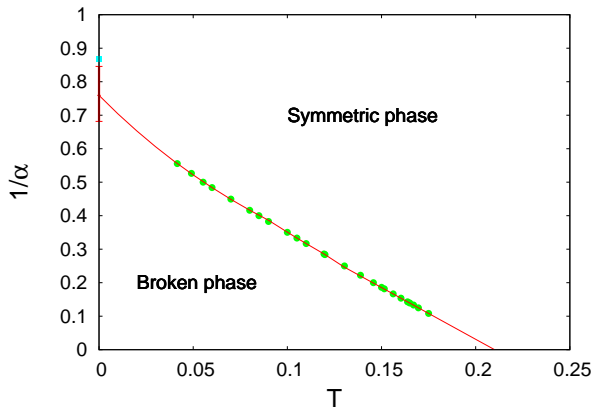}} 
  \caption{The phase boundary curve in the $(T, 1/\alpha)$ plane in the Landau gauge, 
separating the chiral symmetric phase from the broken one.
The error bar assigned to the best fit curve at $T=0$ denotes the error in the
$3\sigma$ accuracy level.}
  \label{phase_diag}
\end{figure}

Now we present Fig.~\ref{phase_diag}, showing the phase boundary curve 
in ($T, 1/\alpha$) plane in the Landau gauge, which separates the chiral symmetric 
phase from the broken one. 
This critical curve shows that the critical coupling inverse $1/\alpha_c$ is a monotonically 
decreasing function of the temperature $T$ slightly concave upwards, and displays two 
characteristic behaviors: (1) as $T$ becomes lower, the critical coupling inverse $1/\alpha_c$ 
becomes larger and seems to increase from below to the zero temperature value $\frac{3}{1.1\pi}$~\cite{Kugo-Naka}, 
and (2) the critical temperature $T_c$ increases as the coupling inverse becomes smaller 
(coupling become stronger), with possible saturation behavior approaching $T_0 \simeq 0.21$ from 
below in the strong coupling limit. 

It is worth noting that the critical coupling at zero temperature is $\alpha^{MS}_c(T=0)=
\frac{1.1}{3} \pi \simeq 1.152$~\cite{Kugo-Naka}, 
and our result predicts slightly larger critical coupling $\alpha_c(T=0) = 1.32 \pm 0.14$.
The errors are given in the $3\sigma$ accuracy level of the least $\chi^2$ fit. 
Phase transition occurs only in the region $\alpha \ge \alpha_c(T=0) \simeq 1.32$ and $T \le T_0 \simeq 0.21$, 
i.e., chiral symmetry broken phase is restricted to the region of the ($T, 1/\alpha$) plane lower than 
the critical curve in Fig.~\ref{phase_diag}. Therefore it is obvious that the region of the coupling and 
temperature where we study the property of the quasifermion is well inside the chiral symmetric phase.

How does the property of the quasifermion change inside the chiral symmetry broken phase?  
This is an interesting question. Does the quasifermion mode still exist in the broken phase? 
These questions will be discussed in a separate paper.

\section{\label{ap:Disp-Law-rho}Dispersion law $\omega^{\rho}(p)$ determined through the
peak position of the spectral density $\rho_+$}

Throughout this paper we determined the dispersion law of the thermal quasifermion with the
on-shell condition Re$[D_+(p_0,p=0)]=0$. As was explained in 
Sec.~\ref{sec:Solution-Dispersion}, generally speaking,
the pole of the propagator or the point where the inverse propagator vanishes defines the
corresponding particle and its dispersion law, and we can use another definition of
on shell. One of such definition is to use the peak position of the spectral density as
the pole position of the corresponding particle, with which we can also determine the
dispersion law of this particle.

These two definitions of on shell almost agree with each other when the imaginary part
of the mass term is small.  In fact, in the weak coupling region at high temperature, the
fermion branch of the dispersion law determined through the peak position of the
spectral density $\rho_+$ almost completely coincides with the dispersion law,
Fig.~\ref{dis_zeros_weak}, determined through the on-shell condition
Re$[D_+(p_0, p=0)]=0$.

There are mainly two reasons why we adopt the on-shell condition Re$[D_+(p_0, p=0)]=0$
rather than that given by the peak position of the spectral density in the present
analysis. The first reason is already explained in Sec.~\ref{sec:Solution-Density-Structure};
it is pointed out there that at the intermediate coupling strength the spectral density exhibits
a typical double peak structure, indicating the existence of two poles in the quasifermion
propagator. This is, however, not the case. There are actually three poles in the propagator in
the corresponding coupling region.  The third peak representing the ultrasoft third pole
is completely hidden under the big tails of the broad two peaks, and thus cannot be observed.
The position of the two peaks does not exactly represent the true position of the pole
either. These facts indicate that information obtained through the analysis of the
spectral density itself is not complete but even misleading.

\begin{figure}[htbp] 
  \centerline{\includegraphics[width=7.5cm,clip]{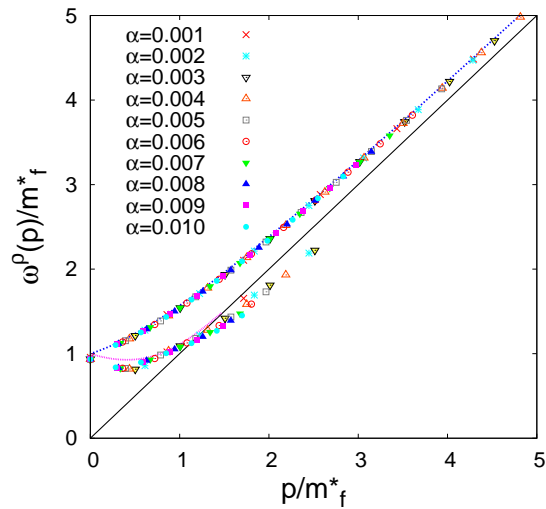}} 
  \caption{The normalized quasifermion dispersion law $\omega^{\rho}_{\pm}/m^*_f$ determined 
through the peak position of the spectral density at $T=0.3$ in the small coupling region. (See text.)}
\label{dis_rho_weak}
\end{figure}

The second reason why did not adopt simply the peak position of the spectral density as the
pole position of the corresponding particle, is the appearance of the plasmino branch
continuing to exist in the spacelike domain, namely the existence of the spacelike
plasmino solution.  In Fig.~\ref{dis_rho_weak} we show the dispersion law $\omega^{\rho}(p)$
determined through the peak position of the spectral density $\rho_+$,
which exactly corresponds to Fig.~\ref{dis_zeros_weak}, showing the dispersion law 
determined through the on-shell condition Re$[D_+(p_0, p=0)]=0$.  Though the fermion
branches almost completely agree with each other, the plasmino branch exhibits a typical 
difference. In Fig.~\ref{dis_zeros_weak}, the plasmino branch exhibits a minimum at $p \neq 0$
and vanishes rapidly on to the light cone as $p$ gets large. In Fig.~\ref{dis_rho_weak}, the
plasmino branch also exhibits a minimum at $p \neq 0$, and approaches rapidly to the
light cone, \textit{then crosses the light cone and continues to exist in the spacelike
domain} of the world sheet.

With these two reasons, in the present analysis we do not adopt defining the on-shell
condition through the peak position of the spectral density.

\begin{acknowledgements}
Part of the present work is supported by the Nara University Research Fund.
We thank the Yukawa Institute for Theoretical Physics at Kyoto University. 
Discussions during the YITP workshop YITP-W-11-14  were useful to complete this work.
Numerical computation in this work was carried out at the Yukawa Institute Computer Facility.
\end{acknowledgements}


\begin{references}
\bibitem{footnote-Decay-width}
    To be exact, in the HTL resummed perturbation calculation, the logarithmic 
factor, $\log (1/g)$, does not appear in the decay width at rest, $\gamma (p=0)$,  
but appears in the decay width of a (fast) moving fermion. In this sense, the present 
result in the DSE analysis does not exactly agree with the results in the HTL resummed 
perturbation calculation~\cite{Nakk-Nie-Pire}.
\end{references}
\end{document}